\documentclass[12pt]{article}
\newlength{\abstractwidth}
\renewcommand{\thefootnote}{\fnsymbol{footnote}}
\renewcommand{\thanks}[1]{\footnote{#1}} 
\newcommand{\starttext}{
\setcounter{footnote}{0}
\renewcommand{\thefootnote}{\arabic{footnote}}}

\newcommand{\be}{\begin{equation}}
\newcommand{\bea}{\begin{eqnarray}}
\newcommand{\eea}{\end{eqnarray}}
\newcommand{\beq}{\begin{equation}}
\newcommand{\ee}{\end{equation}}
\newcommand{\eeq}{\end{equation}}

\def\ba{\begin{eqnarray}}
\def\ea{\end{eqnarray}}

\def\N{{\cal N}}

\def\G{{\cal G}}
\def\L{{\cal L}}

\def\p{{\partial}}

\def\Re{{\rm Re}}

\def\12{{1 \over 2}}
\def\32{{3 \over 2}}
\def\72{{7 \over 2}}
\def\92{{9 \over 2}}

\def\res{{\rm Res}}

\def\P{\Phi}
\def\a{\alpha}
\def\b{\beta}

\def\bL{{\bar\Lambda}}
\def\M{{\cal M}}
\def\Ga{\Gamma}
\def\g{\gamma}
\def\K{{\cal K}}
\def\s{\sigma}

\begin{document}
\baselineskip=16pt

\begin{titlepage}
\bigskip
\hskip 3.7in\vbox{\baselineskip12pt
\hbox{UCLA/02/TEP/29}
\hbox{Columbia/Math/02}}

\bigskip\bigskip\bigskip\bigskip

\centerline{\Large \bf Seiberg-Witten Theory, Symplectic Forms,}
\medskip
\centerline{\Large \bf and Hamiltonian Theory of Solitons
\footnote{Research
supported in part by the National Science Foundation 
under grants PHY-98-19686,
PHY-0140151, DMS-98-00783,
and DMS-01-04621.}}

\bigskip\bigskip
\bigskip\bigskip

\centerline{\bf  Eric D'Hoker$^{a}$, I. Krichever$^{b}$, and
D.H.Phong$^{b}$}

\bigskip
\bigskip

\centerline{$^a$ \it Department of Physics}
\centerline{ \it University of California, Los Angeles, CA 90095}
\bigskip
\centerline{$^b$ \it Department of Mathematics}
\centerline{ \it Columbia University, New York, NY 10027}

\bigskip\bigskip

\begin{abstract}

This is an expanded version of lectures given in Hangzhou and
Beijing, on the symplectic forms common to Seiberg-Witten theory
and the theory of solitons. Methods for evaluating
the prepotential are discussed. The construction
of new integrable models arising from supersymmetric
gauge theories are reviewed,
including twisted Calogero-Moser systems and spin chain
models with twisted monodromy conditions. A practical framework
is presented for evaluating the universal symplectic form
in terms of Lax pairs. A subtle distinction between a Lie algebra
and a Lie group version of this symplectic form is clarified,
which is necessary in chain models.

\end{abstract}
\end{titlepage}

\vfill\eject

\tableofcontents

\vfill\eject

\starttext
\baselineskip=15pt
\setcounter{equation}{0}
\setcounter{footnote}{0}

\section{Introduction}

Soliton equations and integrable models have usually a rich Hamiltonian structure,
in terms of which they become Hamiltonian flows with a full 
set of integrals of motion in involution (see \cite{faddeev}
and references therein). 
However, except for the $R$-matrix 
approach developed by Faddeev and Takhtajan, the 
Hamiltonian structure 
for each model had been found on a case by case basis. In particular,
there was no general construction based directly on the characterizing feature of soliton
equations and integrable models, namely that they can be expressed as a Lax or a zero curvature
equation. Such a construction became available only recently \cite{kp97, kp98}.
A key input came from supersymmetric gauge theories: the Seiberg-Witten Ansatz for their 
exact solution in the Coulomb phase can be expressed in terms of a symplectic form
on a moduli space of spectral curves and divisors \cite{seiberg}. Many spectral curves 
arising in this way were recognized as identical with
the spectral curves of some known integrable models \cite{gorsky, martinec, martinecb, kp97}.
The problem became to identify the symplectic form for the integrable models which would
correspond to the one from the Seiberg-Witten Ansatz. The symplectic form found
in \cite{kp97, kp98} was the answer. It provided at the same time the
direct and universal construction in terms of Lax pairs which had been lacking in the Hamiltonian
theory of solitons.

\medskip
The contents of this paper are as follows.

\medskip
The key information provided by the spectral curves and their symplectic structure is the
prepotential ${\cal F}$. For integrable models, ${\cal F}$ is the $\tau$-function of the 
Whitham hierarchy \cite{kr94}.
For supersymmetric gauge theories, it is the prepotential
which determines entirely the Wilson effective action in the Coulomb phase (see \cite{dp99} and 
references therein). In the perturbative regime, ${\cal F}$ consists of the classical prepotential,
together with one-loop and instanton corrections. The one-loop corrections characterize the
field content of the corresponding supersymmetric gauge theory. They are a defining feature of
the correspondence between gauge theories and integrable models, and several methods for determining
them have been developed \cite{klemm, dkp97a, schnitzer, recent}. Here we present another method
which is efficient, and
based on a $\delta$-regularization process different from the analytic continuation in
\cite{dkp97a, schnitzer}. The instanton corrections can be
recaptured from renormalization group equations \cite{dkp97b}. We provide such
equations in some models where they had not been available \S 2.

\medskip
The correspondence between integrable models and supersymmetric gauge theories
has been mutually beneficial. In one direction, Seiberg-Witten solutions of gauge theiries
have been found via integrable models. Such was the case for gauge theories
with a matter hypermultiplet in the adjoint representation, where the solution
came from twisted Calogero-Moser systems \cite{dp98a, dp98b,dp98c}. In the other direction,
solutions of gauge theories which had been obtained by other methods such
as M theory \cite{witten, LL} or geometric engineering \cite{vafa}, have led to
the discovery of new integrable models
\cite{kp00, kp02}. We review these developments in
\S 4 and \S 5, with emphasis on the scaling limits of Calogero-Moser systems,
and the general construction of spin chain models with twisted monodromy
conditions.

\medskip
The symplectic form of \cite{kp97,kp98} is discussed in Section \S 5.
Given the diversity of integrable models, it may be helpful to provide
a framework which is at the same time broad enough for applications to 
supersymmetric gauge theories, and yet simple enough for the symplectic 
form to be easily worked out. We provide such a framework in \S 5.1. 
In practice, the general construction of \cite{kp97,kp98} can lead 
to two slightly different symplectic forms, depending on whether
the Lax operator is viewed as a Lie group element or as a Lie algebra
element. This distinction is not relevant in most models, but
it is important for chain models such as the Toda chain or the spin models
of Section \S 4. We clarify it in Section \S 5.2.
It is a remarkable fact that the formula for
the finite-dimensional symplectic forms
arising in Seiberg-Witten theory works extends to
the partial
differential equations of soliton theory.
In particular, we obtain a symplectic
structure even for 
zero curvature equations in
$2+1$ space-time variables,
whose Hamilnonian formulation had formerly not
been fully satisfactory. In section \S 5.4,
we present a basic example
of how this can be done, in the case of the Kadomtsev-Petviashvili hierarchy,
following \cite{kp98}. 
In the remaining sections \S 5.3 and \S 5.5, we describe some recent
progress where the general construction of symplectic forms played
a major role. This includes the construction of Lax equations with spectral
parameter on a curve of higher genus, field analogues of Calogero-Moser
equations, and isomonodromy problems \cite{kr01a,kr01b}.   

\section{Seiberg-Witten Solutions of ${\cal N}=2$
Super Yang-Mills Theories}

\subsection{The Wilson effective action}

In four dimensions, ${\cal N}=2$ supersymmetric gauge theories
can be classified by their gauge group $G$ and the representation
$R$ of their matter hypermultiplets, subject to the requirement of
asymptotic freedom or scale invariance. Let $n$ denote the rank of the gauge group $G$.
The ${\cal N}=2$ multiplet of the gauge field $A_\mu dx^{\mu}$
consists of $(A_{\mu}dx^{\mu},\lambda_{\pm},\phi)$,
where $\lambda_{\pm}$ are Weyl spinors, and $\phi$
is a scalar, all valued in the adjoint representation.
Classically, the equations of motion are
\be
F_{\mu\nu}=0,
\ \
D_\mu\phi=0,
\ \
[\phi,\phi^{\dagger}]=0
\ee
where $F_{\mu\nu}$ is the curvature of the gauge field
$A_{\mu}dx^{\mu}$.
Thus the theory admits an $n$-dimensional moduli space of
classical
vacua, corresponding to the diagonalizable elements in the Lie
algebra of $G$. Quantum mechanically, the gauge group is
spontaneously broken
to $U(1)^{n}$, and the effective theory is a theory of
$n$
interacting ${\cal N}=2$ supersymmetric electromagnetic multiplets
$(A_{\mu}^idx^{\mu},\lambda_{\pm}^i,\phi^i)$, $1\leq i\leq\
n$.
In view of the ${\cal N}=2$ supersymmetry, the Wilson low-energy
affective
action ${\cal L}_{eff}$ is characterized completely by a single
function, the
prepotential ${\cal F}(\phi,\Lambda)$
\bea
{\cal L}_{eff}&=&
\big\{{\rm Im}\,\tau_{ij}(\phi)\big\}F_{\mu\nu}^iF^{i\mu\nu}
+
\big\{{\rm Re}\,\tau_{ij}(\phi)\big\}F_{\mu\nu}^i\tilde F^{i\mu\nu}
+\cdots
\nonumber\\
\tau_{ij}&=&{\p^2{\cal F}\over\p\phi^i\p\phi^j}(\phi,\Lambda)
\eea
Here $\tilde F^{i\mu\nu}=
\epsilon^{\mu\nu\lambda\rho}F_{\lambda\rho}^i$, and
$\Lambda$ is a scale introduced by renormalization. In the
perturbative regime where $\Lambda$ is small compared to $\phi$,
the prepotential ${\cal F}$ admits an expansion of the form
\be
\label{log}
{\cal F}(\phi,\Lambda)
=
{i\over 8\pi}
\big(\sum_{\alpha\in {\cal R}(G)}(\alpha\cdot\phi)^2
{\rm ln}\,{(\alpha\cdot\phi)^2\over\Lambda^2}
-
\sum_{\lambda\in {\cal W}(R)}(\lambda\cdot\phi+m)^2
{\rm ln}\,{(\lambda\cdot\phi+m)^2\over\Lambda^2}\big)
+
\sum_{d=1}^{\infty}{\cal F}_d\Lambda^{\big(2h_G^{\vee}-I(R)\big)d}
\ee
where ${\cal R}(G)$ are the roots of $G$,
${\cal W}(R)$ are the weights of the representation $R$,
and $m$ is the mass of the hypermultiplet.
The expression $I(R)$ is the Dynkin index of the
representation $R$. When $R$ is the adjoint representation,
it is also given by $2h_G^{\vee}$, where $h_G^{\vee}$ is the
dual Coxeter number of $G$. In the above expansion, we have ignored the classical prepotential.
The logarithmic singularities are due to one-loop effects,
the higher loops do not contribute by non-renormalization
theorems,
and ${\cal F}_d\Lambda^d$ is the contribution of $d$-instanton
processes. So far we have been discussing asymptotically free
theories.
In scale invariant theories, the renormalization scale $\Lambda$
is replaced by a coupling
\be
\tau={\theta\over 2\pi}+{4\pi i\over g^2}
\ee
which is well-defined microscopically.
Here ${1\over g^2}$ is the gauge coupling and $\theta$ the
$\theta$-angle. The expansion (\ref{log}) for ${\cal F}$
is then replaced by
\be
\label{logtau}
{\cal F}(\phi,\Lambda)
=
{i\over 8\pi}
\big(\sum_{\alpha\in {\cal R}(G)}(\alpha\cdot\phi)^2
{\rm ln}\,(\alpha\cdot\phi)^2
-
\sum_{\lambda\in {\cal W}(R)}(\lambda\cdot\phi+m)^2
{\rm ln}\,(\lambda\cdot\phi+m)^2\big)
+
\sum_{d=1}^{\infty}{\cal F}_d\tau^d
\ee

\subsection{The Seiberg-Witten Ansatz for the effective prepotential}

An exact solution of the gauge theory is provided by the
Seiberg-Witten
Ansatz \cite{seiberg}, which reduces the problem
of finding ${\cal F}$ to finding a fibration of Riemann
surfaces $\Gamma(\Lambda)$ over the moduli space of vacua,
equipped with
a meromorphic $1$-form $d\lambda$ on each surface
$\Gamma(\Lambda)$.
The prepotential ${\cal F}$ is then obtained from
$\Gamma(\Lambda)$
and $d\lambda$ by the Ansatz
\be
\label{sw}
a_i={1\over 2\pi i}\oint_{A_i}d\lambda,
\ \
a_{Di}={1\over 2\pi i}\oint_{B_i}d\lambda,
\ \
a_{Di}={\p{\cal F}\over\p a_i},
\ \
1\leq i\leq n
\ee
where $A_i$, $B_i$ are suitable cycles over the surface
$\Gamma(\Lambda)$.
The fibration $\Gamma(\Lambda)$ will usually include singular
surfaces
over some subvarieties of the moduli space of vacua,
corresponding to when physical massless particles appear.
There are also severe constraints on the fibrations and
differentials
can arise as the Seiberg-Witten solution of a gauge theory.
The curves $\Gamma(\Lambda)$  must be invariant under the Weyl
group
Weyl($G$), which is the residual gauge invariance after the gauge
group $G$ has been broken down to its Cartan subalgebra, $U(1)^n$.
The differential ${\p d\lambda\over \p a_i}$ must be holomorphic,
and the residues of $d\lambda$ must be independent of $a_i$
and linear in the masses $m$. Physically, this means that
the hypermultiplet masses are not renormalized and are consistent
with
the BPS mass formula, in which the hypermultiplet mass parameters
enter linearly.

As noted very early on in \cite{seiberg}, the Seiberg-Witten
solution of a gauge theory can already be derived from a natural
symplectic form
$\omega$ defined by the data $\Gamma(\Lambda),d\lambda$
\be
\omega=\delta\,\big(\sum_{i=1}^{n}d\lambda(z_i)\big)
\ee
where $\delta$ denotes exterior differential on the total space of
the fibration of spectral curves and divisors $[z_1,\cdots,z_n]$. 
This symplectic form can
often be more manageable that the data $(\Gamma(\Lambda),d\lambda$ itself.

\subsection{The logarithmic singularities of the
effective prepotential}

The Seiberg-Witten exact solution is now known for many gauge
theories,
although not for all. Methods for finding the solution include
geometric engineering \cite{vafa}, M theory \cite{witten, LL}, and integrable
models
\cite{martinec, donagi, dp98a, dp98b, dp98c, marshakov},
the latter being the main one considered here. Once certain models
have been solved, the solution of others can also be derived by
various
decoupling limits. In all cases, it is a key requirement that
the fibration $\Gamma(\Lambda)^{n}$ for a gauge group
$G$ and a representation $R$ of the matter hypermultiplet
must exhibit the corresponding logarithmic singularities
(\ref{log}) or (\ref{logtau}) in the perturbative regime of large
vacuum expectation values. The problem of identifying the
logarithmic singulariies
(as well as the instanton corrections) from a given fibration
$(\Gamma(\Lambda),d\lambda)$ has thus received considerable
attention. Some of the many methods developed are Picard-Fuchs
equations
\cite{klemm}, the method of residues \cite{dkp97a},
non-hyperelliptic extensions of the method of residues
\cite{schnitzer},
and others \cite{recent, whitham, wdvv}.
We take the opportunity in this paper to present a new method
which may also be useful.
To be specific, we shall consider the case of $G=SU(N)$,
with either no hypermultiplet, or a hypermultiplet in the
antisymmetric
or the symmetric representation. The integrable models
corresponding to
these theories are also the focus of \S 4 below.

\subsubsection{The pure $SU(N)$ Yang-Mills theory}

As a warm-up, we begin with the case of pure Yang-Mills. Suitably
formulated,
we shall see that the more difficult cases of a symmetric or an
anti-symmetric
hypermultiplet follow from the present method by simple
modifications.
For the pure $SU(N)$ Yang-Mills theory, the following Seiberg-Witten
differentials and differential were proposed
(see \cite{su(n)} and references therein)
\be
\label{toda}
{1\over 2}(k+{\Lambda^N\over k})=P(x),
\ \ d\lambda=x\,d\,{\rm ln}\,k
\ee
where $P(x)$ is a monic polynomial of degree $N$ with no
$u_{N-1}x^{N-1}$ term, whose $N-1$ coefficients can be viewed as parameters
for the moduli of classical vacua. For the derivation of the
effective prepotential, it is more convenient to introduce the
variable $y=k-P(x)$, and to parametrize $P(x)$ as
$P(x)=\prod_{k=1}^N(x-\bar a_k)$, so that the periods $a_k$ of $d\lambda$ will 
emerge naturally as renormalizations
of the classical moduli parameters $\bar a_k$. With these variables,
the curve and differential are now given by
\be
y^2=P(x)^2-\Lambda^{2N},
\ \ \
d\lambda=x{dP\over y}.
\ee
This is a hyperelliptic curve, made of two copies of the complex
plane, glued along $N$ cuts going between pairs of zeroes of the
right hand side. More specifically, let $x_k^{\pm}$ be the zeroes of 
the right hand side, with
$x_k^{\pm}\to\bar a_k$ as $\Lambda\to 0$. 
For real values of $\bar a_k$ and $\bL$, we let $x_k^{\pm}$
be respectively the left and right edges of the cuts.
We choose the cycles $A_k$
to be loops around the cuts from $x_k^-$ to $x_k^+$, and the
cycles $B_k$ to be the cycles going from $x_1^+$ to $x_k^-$ on
one sheet, and coming back on the other sheet. Our first task is to determine the 
logarithmic singularities of $a_{Dk}$
\be
\label{bperiods}
a_{Dk}={1\over 2\pi i}\oint_{B_k}d\lambda
=
{1\over \pi i}\int_{x_1^+}^{x_k^-}x{dP\over \sqrt{P^2-\bar\Lambda^2}},
\ \ \
2\leq k\leq N
\ee
Here we have set $\bar\Lambda=\Lambda^N$. Locally, $a_{Dk}$
is a holomorphic function of $\bL\not=0$. It is a multivalued
function, due to the choice of branch points $x_1^+$ and $x_k^-$.
If we analytically continue from $\bL^2$ to $e^{2\pi i}\bL^2$,
the $B_k$ cycle transforms to $B_k+A_k-A_1$. Thus
\be
2\pi i\,a_{Dk}=(\bar a_k-\bar a_1)\ln \bL^2+b(\bL)
\ee
where $b(\bL)$ is a {\it single valued} holomorphic function
on the punctured disk $\bL\not=0$. Our next goal is to show that
$b(\bL)$ is {\it bounded}.
This will imply that $b(\bL)$ is a holomorphic function of $\bL$
on the whole disk. Fix $\delta$ small and rewrite (\ref{bperiods})
as the sum of three integrals
\be
\label{integral}
\pi i\,a_{Dk}
=
\int_{\bar a_1+\delta}^{\bar a_k-\delta}x{dP\over
\sqrt{P^2-\Lambda^2}}
+
\int_{\bar a_k-\delta}^{x_k^-}x{dP\over \sqrt{P^2-\Lambda^2}}
-
\int_{\bar a_1+\delta}^{x_1^+}x{dP\over \sqrt{P^2-\Lambda^2}}
\ee
The first integral $I_1$ is a holomorphic function of
$\bL$ for $|\bL|<<\delta$. Therefore
\be
I_1=I_1^0+{\cal O}(\bL)
\ee
where
\bea
I_1^0
&=&
\int_{\bar a_1+\delta}^{\bar a_k-\delta}
x{dP\over P}
=
\int_{\bar a_1+\delta}^{\bar a_k-\delta}
\sum_{j=1}^N(1+{\bar a_j\over (x-\bar a_j)})dx
\\
&=&
\bar a_k(1+\ln(-\delta))-\bar a_1(1+\ln \delta)
+
\sum_{j\not=k}\bar a_j\ln\,(\bar a_k-\bar a_j)
-
\sum_{j\not=1}\bar a_j\ln (\bar a_1-\bar a_j)
+{\cal O}(\delta)\nonumber
\eea
The second integral in (\ref{integral}) is equal to
\be
I_2
=
\bar a_k\int_{\bar a_k-\delta}^{x_k^-} {dP\over
\sqrt{P^2-\Lambda^2}}
+
\int_{\bar a_k-\delta}^{x_k^-} {(x-\bar a_k)\,dP\over
\sqrt{P^2-\Lambda^2}}
\ee
In the range of integration, we have
$(x-\bar a_k)=\prod_{j\not=k}(\bar a_k-\bar a_j)^{-1}P(x)(1+{\cal
O}(\delta))$. The term ${\cal O}(\delta)$ is readily
seen to contribute only another ${\cal O}(\delta)$ term
to $I_2$. As for the other term, an explicit calculation gives
\be
\int_{\bar a_k-\delta}^{x_k^-} {(x-\bar a_k)\,dP\over
\sqrt{P^2-\Lambda^2}}
=
{1\over \prod_{j\not=k}(\bar a_k-\bar a_j)}
\bigg[\sqrt{P^2-\bL^2}\bigg]_{\bar a_k-\delta}^{x_k^-}
=
{\cal O}(\delta)
\ee
In this last equality, we have used the equations
$P^2(x_k^-)=\bL^2$, $P(\bar a_k-\delta)={\cal O}(\delta)$.
To evaluate the remaining term in $I_2$, let
$u$ be the new variable defined by
$P={1\over 2}\bL(e^u+e^{-u})$.
Then
\be
\int_{\bar a_k-\delta}^{x_k^-} {dP\over \sqrt{P^2-\Lambda^2}}
=
\int_{u_k}^0du=-u_k
\ee
where the lower bound of integration is defined by
\be
e^{u_k}
=
{P(a_k-\delta)+\sqrt{P^2(\bar a_k-\delta)-\bL^2}
\over \bL}={2P(\bar a_k-\delta)\over\bL}+{\cal O}(\delta)
\ee
Thus
\be
I_2=-\bar a_k\ln
\big({2(-\delta)\prod_{j\not=k}(\bar a_k-\bar a_j)\over\bL}
\big)+{\cal O}(\delta)
\ee
The third term in (\ref{integral}) admits a similar expression.
Altogether, we find
\be
\pi i\,a_{Dk}
=
(\bar a_k-\bar a_1)
(\ln \bL+1-\ln 2)
-
\bigg(\sum_{j\not=k}(\bar a_k-\bar a_j)\ln (\bar a_k-\bar a_j)
-
(k\to 1)\bigg)
+
{\cal O}(\delta)
\ee
This equation implies that $b(\bL)$ is bounded, and hence
holomorphic.
Now $b(\bL)$ does not depend on $\delta$. Thus its leading term
can be obtained by letting $\delta\to 0$. Taking into account
the fact that $a_k=\bar a_k+{\cal O}(\Lambda)$,
we conclude that
\be
\pi i \,a_{Dk}
=
(a_k-a_1)(\ln\,\bL+1-\ln\,2)
-
\bigg(\sum_{j\not=k}(a_k-a_j)\ln(a_k-a_j)
-
(k\to 1)\bigg)
+
{\cal O}(\bL)
\ee
These are indeed the logarithmic singularities 
of the effective
prepotential characteristic of the pure $SU(N)$ 
Yang-Mills theory.

\medskip

The effective prepotential ${\cal F}$ of supersymmetric gauge theories
has a lot of structure. Of particular interest to us is a renormalization
group equation satisfied by ${\cal F}$. This renormalization group equation 
provides another
important link with a Hamiltonian formulation of integrable models,
and is also an efficient tool for the evaluation of instanton
corrections. It is a consequence
of a closed formula for ${\cal F}$. This formula goes back
to a similar formula for the
$\tau$-function of the Whitham hierarchy \cite{kr94}, 
and can be expressed as follows in the case of pure $SU(N)$ Yang-Mills
\be
2{\cal F}={1\over 2\pi i}\bigg(\sum_{k=2}^Na_k\oint_{B_k}d\lambda
+
{\rm Res}_{P_+}(xd\lambda){\rm Res}_{P_+}(x^{-1}d\lambda)
+
{\rm Res}_{P_-}(xd\lambda){\rm Res}_{P_-}(x^{-1}d\lambda)
\bigg)
\ee
Here $P_{\pm}$ are the two points on $\Gamma$ above $x=\infty$.
To establish this formula, it suffices to show that the derivative
of the right hand side with respect to $a_m$ is equal to $2a_{Dm}$.
Now the residues of $d\lambda$ as well as
the residues of $x^{-1}d\lambda$ at $P_{\pm}$ are fixed. 
If we denote by $d\omega_k$ the basis of holomorphic differentials dual to $A_k$, it follows that 
\be
{\partial\over\partial a_m}d\lambda=2\pi i d\omega_m,
\ee
and hence the derivative of the above right hand side 
with respect to $a_m$ is given by
\be
{1\over 2\pi i}\oint_{B_m}d\lambda
+\sum_{k=2}^Na_k\oint_{B_k}\,d\omega_m
+
{1\over 2\pi i}\big({\rm Res}_{P_+}(x^{-1}d\lambda){\rm Res}_{P_+}(xd\omega_m)
+
{\rm Res}_{P_-}(x^{-1}d\lambda){\rm Res}_{P_-}(xd\omega_m)
\big)
\ee
Let $d\Omega_{\pm}$ be the Abelian differentials
of the second kind with a double pole at $P_\pm$ respectively,
vanishing $A_k$-periods, and normalization $d\Omega_{\pm}=dx\big(1+{\cal O}(x)\big)$. 
By the Riemann bilinear relations, 
\be
\oint_{B_k}d\omega_m=\oint_{B_m}d\omega_k,
\ \ \
{\rm Res}_{P_{\pm}}(xd\omega_m)
=
{1\over 2\pi i}\oint_{B_m}d\Omega_{\pm}
\ee
we can rewrite the last three terms on the above right hand side in
terms of integrals over the cycle $B_m$.
The resulting form is precisely $d\lambda$
\be
d\lambda=\sum_{k=2}^N2\pi i\,a_kd\omega_k
+{\rm Res}_{P_+}(x^{-1}d\lambda)\,d\Omega_+
+{\rm Res}_{P_-}(x^{-1}d\lambda)\,d\Omega_-
\ee
as an inspection of the poles, residues, and periods of $d\lambda$
readily shows. The closed formula for ${\cal F}$ can also be 
rewritten as
\bea
\big(\sum_{k=2}^Na_k{\partial\over\partial a_k}-2\big){\cal F}  
&=&
-{1\over 2\pi i}
\big(
{\rm Res}_{P_+}(xd\lambda){\rm Res}_{P_+}(x^{-1}d\lambda)
+
{\rm Res}_{P_-}(xd\lambda){\rm Res}_{P_-}(x^{-1}d\lambda)\big)
\nonumber\\
&=&
-{N\over 2\pi i}\,\sum_{k<m}\bar a_k\bar a_m.
\eea
By dimensional analysis, we have $(\Lambda{\partial\over \partial\Lambda}
+\sum_{k=2}^Na_k{\partial\over\partial a_k}-2){\cal F}=0$. Thus we obtain
the following renormalization group equation, giving the
dependence of the effective prepotential on the renormalization scale $\Lambda$
\cite{dkp97c}
\be
\Lambda{\partial\over \partial\Lambda}{\cal F}
=
{N\over 2\pi i}\,\sum_{k<m}\bar a_k\bar a_m.
\ee
The renormalization group equation provides no information on
the logarithmic singularities of the effective prepotential,
which have to be determined independently as we did above,
in order to verify that the Seiberg-Witten curve proposed
are indeed the solution of a given gauge theory.
But it gives a relatively easy way of determining the instanton
corrections to any order, since the terms $\bar a_k$, and hence the
``beta function" ${N\over 2\pi i}\,\sum_{k<m}\bar a_k\bar a_m$,
can be readily evaluated perturbatively in terms of the periods $a_k$.

\medskip

We shall see later that the Seiberg-Witten curve (\ref{toda})
coincides with the spectral curve for the $SU(N)$ affine Toda system.
Furthermore, the Toda system can be given a Hamiltonian formulation
with Hamiltonian given precisely by the above right hand side.
It had been pointed out by Donagi and Witten \cite{donagi} that the
Seiberg-Witten Ansatz provides a symplectic structure with respect to
which the moduli vacua parameters become a maximal system of
commuting Hamiltonians. The renormalization group equation
suggests a more precise picture: the beta function 
is the Hamiltonian of a very specific integrable model, whose spectral curve coincides 
with the Seiberg-Witten curve of the gauge theory.

\subsubsection{The $SU(N)$ Yang-Mills theory with antisymmetric
matter}

We come now to the derivation of the logarithmic singularities for
the effective prepotential from the
Landsteiner-Lopez curves for the gauge theory with a
hypermultiplet in the antisymmetric representation.
This had been carried out by Ennes, Naculich, Rhedin, and
Schnitzer \cite{schnitzer}, by extending to the
non-hyperelliptic case the method of residues introduced in
\cite{dkp97a}. Here we shall derive these
singularities by the $\delta$ regularization method used in the
previous section.

\medskip

The Seiberg-Witten curve for the $SU(N)$ theory with a matter
hypermultiplet in the antisymmetric representation
had been proposed by Landsteiner-Lopez, using a brane construction
\cite{LL}. It is of the form
\be
\label{LL}
\Gamma\ :\ y^3-(3\Lambda^{N+2}+x^2P(x))y^2+
(3\Lambda^{N+2}+x^2P(-x))\Lambda^{N+2}y-
\Lambda^{3(N+2)}=0, \label{ll0}
\ee
where $\Lambda$ is the renormalization scale, and $P(x)$
is again a monic polynomial of degree $N$, without $u_{N-1}x^{N-1}$ term.
The Seiberg-Witten differential $d\lambda$ is given by
\be
d\lambda=x{dy\over y}. \label{sw1}
\ee
We set $P(x)=\sum_{i=0}^Nu_ix^i=\prod_{k=1}^N(x-\bar a_k)$, $u_N=1$, $u_{N-1}=0$.
The curve $\Gamma$ is a three-sheeted cover of the complex plane.
It is invariant with respect to the involution
\be
\label{2}
\sigma: y\to y^{-1}\Lambda^{2(N+2)},\ x\to -x\ .
\ee
The quotient $\Gamma_0=\Gamma/\sigma$ has genus $N-1$. 
We denote the three points on $\Gamma$ above $x=\infty$ by
$P_1,P_2,P_3$, with $P_1$ characterized by $y\sim x^{N+2}$,
$P_3=\sigma(P_1)$, and $P_2=\sigma(P_2)$.

\medskip
We consider now the $\Lambda\to 0$ limit.
Set $\bL^2=\Lambda^{(N+2)}$. Then the
three branches
$y_i(x),\ i=1,2,3$ of the spectral curve
(\ref{LL}) in the limit $\Lambda\to 0$ can be obtained in the form
of series in $\bL^2$:
\beq\label{l1}
y_1=x^2P(x)+{3P(x)-P(-x)\over
P(x)}\bL^2+\sum_{s=2}^{\infty}\xi_{1,s}(x)
\bL^{2s}
\eeq
\beq\label{l2}
y_2=\bL^2\left({P(-x)\over P(x)}+{(P(-x)-P(x))^3\over
P^3(x)P(-x)}\bL^2+
\sum_{s=2}^{\infty}\xi_{2,s}(x)\bL^{2s}\right)
\eeq
\beq\label{l3}
y_3=\bL^4{1\over y_1(-x)}=\bL^4\left({1\over
x^2P(-x)}+{\cal 0}(\bL^2)\right)
\eeq
The first series is convergent outside the neighborhood of the zeroes
of $x^2P$ where $|\bL^2/P(x)|<c_1$ for some constant $c_1$. The second series is 
convergent outside neighborhoods of the zeroes of $x^2P(x)$ and
$x^2P(-x)$.

The branch points of the spectral curve are defined by the equation
of the curve together with the equation
\beq\label{l4}
3y^2-2(3\bL^2+x^2P(x))y+\bL^2(3\bL^2+x^2P(-x))=0.
\eeq
A set of $2N$ solutions of 
these equations can be
found in the form
of series in $\bL$ with the leading terms defined by
\beq\label{l5}
y={1\over 2}x^2P(x)+{\cal O}(\bL^2),\ \ x^2P^2(x)=\bL^2P(-x)+{\cal O}(\bL^3).
\eeq
The corresponding branch points $x_k^{\pm}$ have the form
\beq\label{l6}
x_k^{\pm}=\bar a_k\pm \bL{(-1)^{N/2} \prod_{j}(\bar a_k+\bar a_j)^{1/2}\over
\bar a_k\prod_{j\neq k}(\bar a_k-\bar a_j)}+{\cal O}(\bL^2).
\eeq
They are branch points for the first and the second sheets.
Another $2N$ branching points are $-x^{\pm}_k$. Finally, there are
two more branch points near $x=0$.

Let the $B_k$ -cycle on $\Gamma_0$ be covered by the cycle on $\Gamma$ which goes from
$x_1^+$ to $x_k^-$ on the first sheet and returns back on the
second sheet. Then
\beq\label{s1}
2\pi i\, a_{Dk}=\oint_{B_k}d\lambda=\int_{x_1^+}^{x_k^-}x{dy_1\over y_1}-
\int_{x_1^+}^{x_k^-}x{dy_2\over y_2},\ \ k=2,\ldots N.
\eeq
We follow now closely the discussion of the case of pure
Yang-Mills.
The above integral is a multivalued holomorphic function of
$\bL\neq 0$, due
to the choice of branch points
$x_1^+$ and $x_k^-$. For real values of $\a_k$ and $\bL$ we choose
$x_k^{\pm}$ to be the left and the right edges of cuts.
The path from $\bL^2$ to $e^{2\pi i}\bL^2$ generates a transformation
of the $B_k$-period from $B_k$ to $B_k+A_k-A_1$. Therefore,
\beq\label{s2}
2\pi i\,a_{Dk}=(\bar a_k-\bar a_1)\ln \bL+b(\bL),
\eeq
where $b(\bL)$ is a single-valued holomorphic function in
the punctured disc $\bL\neq 0$. As before, our main goal is to show
that $b(\bL)$ is bounded. That will imply that $b(\bL)$
is a holomorphic function of $\bL$.
Let us again fix $\delta$ and rewrite (\ref{s1}) as a sum of three
terms
\beq
2\pi i\,a_{Dk}=I_1+I_2+I_3,
\eeq
where
\beq\label{s4}
I_1=\int_{\bar a_1+\delta}^{\bar a_k-\delta}x{dy_1\over y_1}-
\int_{\bar a_1+\delta}^{\bar a_k-\delta}x{dy_2\over y_2}\ ,
\eeq
\beq \label{s45}
I_2=\int_{\bar a_k-\delta}^{x_k^-}x{dy_1\over y_1}-
\int_{\bar a_k-\delta}^{x_k^-}x{dy_2\over y_2}\ ,
\eeq
\beq
I_3=\int_{\bar a_1+\delta}^{x_1^+}x{dy_1\over y_1}-
\int_{\bar a_1+\delta}^{x_1^+}x{dy_2\over y_2}\ .
\eeq
In the first integral $I_1$, for $|\bL|<<\delta$, we can replace $y_1,
y_2$ by their leading terms in the expansions (\ref{l1}, \ref{l2}).
Therefore,
\begin{eqnarray}
I_1&=&2\int_{\bar a_1+\delta}^{\bar a_k-\delta}xd\ln (x^2P(x))-
\int_{\bar a_1+\delta}^{\bar a_k-\delta}xd\ln
(x^2P(-x))=(N+2)(\bar a_k-\bar a_1)
\nonumber \\
&&+2\left(\bar a_k\ln(-\delta)-\bar a_1\ln \delta+
\sum_{j\neq k} \bar a_j\ln(\bar a_k-\bar a_j)
-\sum_{j\neq 1} \bar a_j\ln(\bar a_1-\bar a_j)\right)\nonumber \\
{}&&
-\sum_{j}\bar a_j\ln\left({\bar a_k+\bar a_j\over \bar a_1+\bar a_j}\right)+
{\cal O}(\delta) \label{s6}
\end{eqnarray}
Now let us consider the second term.
The equations (\ref{l5}) imply that
$|y(x_k^{\pm})|>c|\bL|$. The constant $c$ can be chosen so that
the inequality
$|y|>c|\bL|$
holds for all points of the path of integration in (\ref{s45}).
The equation (\ref{LL}) implies that
\beq\label{l8}
x^2P(x)y^{-1}=1
-3\Lambda^{N+2}y^{-1}+
(3\Lambda^{N+2}+x^2P_N(-x))\Lambda^{N+2}y^{-2}-\Lambda^{3(N+2)}y^{-3}
\eeq
Therefore $x^2P(x)y^{-1}$ is uniformly bounded
along the path of integration in (\ref{s45}). At the same time,
we have $(x-\bar a_k)=\prod_{j\neq k}\bar a_k^{-2}(\bar a_k-\bar a_j)^{-1}P(x)
\big(1+{\cal O}(\delta)\big)$. Hence,
\begin{eqnarray} \label{s46}
I_2&=&\bar a_k\left(\int_{\bar a_k-\delta}^{x_k^-}{dy_1\over y_1}-
\int_{\bar a_k-\delta}^{x_k^-}{dy_2\over y_2}\right)+{\cal O}(\delta)\nonumber\\
{}&=&\bar a_k\left(\ln y_2(\bar a_k-\delta)-\ln
y_1(\bar a_k-\delta)\right)+{\cal O}(\delta).
\end{eqnarray}
The expansions (\ref{l1},\ref{l2}) for $y_i$ imply
\beq\label{s7}
I_2=\bar a_k\left[2\ln \bL-2\ln (-\delta)-
\ln \left ({\bar a_k^2\prod_{j\neq k}(\bar a_k-\bar a_j)^2\over
(-1)^N\prod_j (\bar a_k+\bar a_j)}\right)\right]+{\cal O}(\delta).
\eeq
A similar expresssion can be derived for $I_3$.
Altogether, we find
\begin{eqnarray}
\label{s10}
2\pi i \,a_{kD}&=&
\bar a_k(2\ln\,2 \bL+(N+2)+N\ln (-1))\nonumber\\
{}&&-\sum_{j\neq k} (\bar a_k-\bar a_j)\ln(\bar a_k-\bar a_j)^2+
\sum_{j<k} (\bar a_k+\bar a_j)\ln(\bar a_k+\bar a_j)^2\nonumber\\
&&-(k\to 1)+{\cal O}(\delta)
\end{eqnarray}
This equation implies that $b(\bL)$ in (\ref{s2}) is uniformly
bounded.
Therefore it is holomorphic. 
Since $b(\bL)$ does not depend on $\delta$,
its leading term can be obtained by letting
$\delta\to 0$. Taking into account 
the fact that $a_k=\a_k+O(\Lambda)$
we conclude that
\begin{eqnarray}\label{s20}
2\pi i \,a_{kD}&=&
a_k(2\ln\,2 \bL+(N+2)+N\ln (-1)) \nonumber\\
{}&&-\sum_{j\neq k}(a_k-a_j)\ln(a_k-a_j)^2+
\sum_{j<k} (a_k+a_j)\ln(a_k+a_j)^2\nonumber\\
&&-(k\to 1)+{\cal O}(\bL)
\end{eqnarray}
Since the weights of the antisymmetric representation are
$e_j+e_k$, $j<k$, these are the logarithmic singularities
of the $SU(N)$ theory with matter in the antisymmetric representation.

\medskip

We derive now the renormalization group equation for this model.
This is a consequence of the following closed formula for the prepotential
\beq
\label{F}
2{\cal F}={1\over 2\pi i}\left[\sum_{k=1}a_k\oint_{B_k^0} d\lambda+
(N+2)\res_{P_+}(xd\lambda)\right]\ .
\eeq
Recall that
$B_k^0$ is a $B$-cycle on $\Gamma_0$,
which is covered twice by the $B_k$ cycle in
$\Gamma$, and $P_+$ is the image of $P_1$ and $P_3$. This can be proved as in the
previous section by differentiating the right hand side with respect to $a_k$, 
applying the Riemann bilinear relations,
and recognizing $d\lambda$ as
\beq\label{4}
d\lambda=(N+2)d\Omega_2^++\sum_{k=1}^{N-1}2\pi i a_kd\omega_k.
\eeq
Here $d\Omega_2^+$ is an {\it even} normalized differential of the
second kind with poles of the form
\beq\label{5}
d\Omega_2^+=\pm dx(1+O(x^{-2}))
\eeq
at the punctures $P_1$ and $P_3$. This differential can also be
considered
as a differential on $\Gamma_0=\Gamma/\sigma$ with a pole at a point
$P_+$, which is the projection of the points $P_1$ and $P_3$.
Similarly, the differentials $d\omega_k$ are normalized {\it even}
differentials on
$\Gamma$ and therefore are
preimages of the basic holomorphic differentials on $\Gamma_0$.
The residue ${\rm Res}_{P_+}(xd\lambda)$
can be easily calculated
\beq
\res_{P_1}(xd\lambda)=-\res_{P_3}(xd\lambda)=
\res_{P_+}(xd\lambda)=
2u_{N-2}
\eeq
In (\ref{F}) we considered $d\lambda$ as a
differential on $\Gamma_0$. If we
lift it to $\Gamma$, ${\cal F}$ can be rewritten as
\beq\label{F1}
2{\cal F}={1\over 4\pi i}\left[\sum_{k=1}a_k\oint_{B_k^+} d\lambda+
(N+2)\res_{P_1}(xd\lambda)-(N+2)\res_{P_3}(xd\lambda)\right].
\eeq
The closed formula for ${\cal F}$ can be converted as before
into a renormalization group equation using the homogeneity
relation $(\Lambda\p_{\Lambda}+\sum_{k=2}^Na_k\partial_{a_k}-2){\cal F}=0$. The result is
\be
\Lambda{\partial\over\partial\Lambda}{\cal F}
={N+2\over \pi i}u_{N-2}
\ee
As noted before, the renormalization group
equation can be used very effectively for the explicit 
evaluation of the contributions
${\cal F}_d$ of instanton processes.

\subsubsection{The $SU(N)$ Yang-Mills theory with symmetric matter}

For the $SU(N)$ theory with matter in the symmetric
representation,
Landsteiner and Lopez \cite{LL} have proposed the following spectral curve
and differential
\bea
\label{LLs}
&&
\Gamma\ :\ y^3+P(x)y^2+P(-x)x^2\Lambda^{N-2}y+\Lambda^{3(N-2)}x^6=0
\nonumber\\
&&
d\lambda=x{dy\over y},
\eea
where $P(x)$ is a polynomial as in the previous model,
that is, it is monic, of degree $N$, with no $u_{N-1}x^{N-1}$ term.
The weights for the symmetric representation are $e_i+e_j$, $i\leq j$,
and the one-loop correction to the effective prepotential must then
be of the form
\be
{1\over 8\pi i}
\big(\sum_{j\not=k}(a_j-a_k)^2\ln \,{(a_j-a_k)^2\over\Lambda^2}
+
\sum_{j\leq k}(a_j+a_k)^2\ln \,{(a_j+a_k)^2\over\Lambda^2}\big)
\ee
These logarithmic singularities can be derived from the proposed
curve and differential in complete analogy with the previous models.
We shall not give the details, but just the closed formula for
the prepotential ${\cal F}$ and the corresponding renormalization
group equation. As in the case of a hypermultiplet in the antisymmetric
representation, the curve is a three-sheeted covering
of the complex plane,
with an involution $\sigma$ given this time by
\be
\sigma: (x,y) \to (-x,{\Lambda^{2(N-2)}x^4\over y})
\ee
Let $P_1$ be the point above $x=\infty$ corresponding to $y\sim 
-x^N$, $P_3=\sigma(P_1)$, and let the differentials $d\Omega_{\pm}$
be the even and odd Abelian differentials with double poles at $P_1$
and $P_3$, with the same normalizations as before (\ref{5}).
Then we have
\be
d\lambda
=
(N-2)\,d\Omega_+
+
2\, d\Omega_-
+
\sum_{k=2}^N2\pi i\,a_kd\omega_k
\ee
from which the desired formula for ${\cal F}$ follows
\be
2{\cal F}
=
{1\over 2\pi i}
\bigg(\sum_{k=2}^Na_k\oint_{B_k}d\lambda
+
(4-N)\,\res_{P_1}(xd\lambda)
+
N\,\res_{P_3}(xd\lambda)\bigg)
\ee
Evaluating the residues and expressing the equation
in terms of $\Lambda$ derivatives, we obtain the following
renormalization group equation
\be
\Lambda{\partial\over\partial\Lambda}{\cal F}
={N-2\over \pi i}u_{N-2}
\ee
As could have been anticipated on general grounds, the coefficient of 
the beta function in each case is proportional to
to $2h_G^{\vee}-I(R)$ (the dual Coxeter number for $SU(N)$ is 
$h_{SU(N)}^{\vee}=N$, and the Dynkin index $I(R)$ is respectively
$N-2$ and $N+2$ for the antisymmetric and the symmetric representations).
A priori, there does not appear to be a reason
for the remaining term $u_{N-2}$ in the beta function
to be the Hamiltonian of an integrable dynamical system. 
This will however be shown in sections \S 3 and \S 4.

\section{Hypermultiplets in the Adjoint Representation 
and Twisted Calogero-Moser Systems}

\setcounter{equation}{0}

\subsection{The ${\cal N}=2$ SUSY Yang-Mills theory with adjoint
hypermultiplet}

In this section, we describe the solution of a class of supersymmetric
gauge theories where the relation with integrable models
plays a major role. These are the
gauge theories with arbitrary simple gauge algebra $G$ and a matter hypermultiplet in the adjoint representation of $G$.
Physically, they are of particular importance as 
scale invariant theories which admit a well-defined microscopic coupling 
$\tau={\theta\over 2\pi}+{4\pi i\over g^2}$. They also admit scaling limits to many other theories
of interest. These properties provide valuable clues
in their eventual solution. First, the parameter $\tau$
must figure in the Seiberg-Witten curves, and in view of Montonen-Olive duality, it is natural
to expect that the Riemann surfaces $\Gamma(\tau)$ should be
coverings of the torus ${\bf C}/{\bf Z}+\tau{\bf Z}$. Second, 
in the limit when the hypermultiplet becomes infinitely massive
and decouples,
the Seiberg-Witten solution of the theories should
reduce to that of the pure Yang-Mills theory. Finally, as the
hypermultiplet mass tends to $0$, the theories acquire an ${\cal
N}=4$ symmetry, and the classical prepotential should receive no
corrections. We shall see how these considerations lead to
the spectral curves of a new integrable model, namely
twisted Calogero-Moser systems \cite{dp98a, dp98b, dp98c}.

\subsection{The $SU(N)$ theory and Hitchin Systems}

The starting point is the solution first found by
Donagi and Witten \cite{donagi} in the basic case where
the gauge group is $SU(N)$.
The key to their construction is a $SU(N)$ Hitchin system,
consisting of a $SU(N)$ connection
coupled to a Higgs field $\phi$ on the torus ${\bf C}/{\bf Z}+\tau{\bf Z}$ with a pole 
with given residue matrix at $0$
\footnote{Strictly speaking, $\phi(z)$ also has an essential singularity
at $0$, which can be gauged away by a singular gauge transformation.
Alternatively, $\phi(z)$ is a meromorphic section, without essential
singularity, of a suitable vector bundle on the torus}.
Let the matrix $\mu$ be defined by
\be
\mu=\pmatrix{1 &\cdot&\cdot\cdot & 0 &0 \cr
\cdot&\cdot&\cdot&\cdot&\cdot\cr
:&\cdot&1 &0&0\cr
0&\cdot&0 &1&0\cr
0&\cdot&0 &0&-(N-1)\cr}
\ee
Then the moduli space 
\be
X_{\mu}
=
\{(A,\phi);\bar\partial_A\phi=m\,\mu\,\delta(z,0),
\ z\in {\bf C}/{\bf Z}+\tau{\bf Z}\}/\{Gauge\ transformations\},
\ee
has the same dimension $N-1$ as the space of vacua of the
four-dimensional gauge theory. The Seiberg-Witten exact solution
for
the $SU(N)$ theory with adjoint hypermultiplet is given by
\be
\Gamma=\{(k,z);\ det\big(kI-\phi(z)\big)=0\},\ \ \
d\lambda=kdz
\ee
By construction, it satisfies the Montonen-Olive $SL(2,{\bf Z})$
duality on $\tau$.
It also satisfies the key scaling consistency checks
described above. As
$m\to 0$, the natural symplectic form on $X_{\mu}$ reduces to
the uncoupled form $\omega=\sum{dx\over y}\wedge da$ on $N-1$
copies
of the torus. As
$m\to\infty$, $\Gamma$ scales to $y^2=\prod_{k=1}^N(x-\bar
a_k)^2-\Lambda^{2N}$,
which is the Seiberg-Witten curve for the
pure $SU(N)$ Yang-Mills theory that we had encountered before in
\S 2.3.1.

\subsection{The $SU(N)$ theory and Calogero-Moser systems}

To see how twisted Calogero-Moser systems emerge
in the solution of the general ${\cal N}=2$ SUSY
gauge theory with adjoint hypermultiplet and gauge group an
arbitrary simple Lie group $G$, 
we need to reexamine the Donagi-Witten solution for
$SU(N)$ in
terms of $SU(N)$ Calogero-Moser systems. The $SU(N)$
Calogero-Moser
system is the Hamiltonian system defined by
\be
H_{CM}^{SU(N)}(x,p)
=
{1\over 2}\sum_{i=1}^Np_i^2
-{1\over 2}m^2\sum_{j\not=k}\wp(x_j-x_k)
\ee
The basic property of the $SU(N)$
Calogero-Moser system
is the existence of a Lax pair $L(z),M(z)$ with spectral
parameter $z\in {\bf C}/{\bf Z}+\tau{\bf Z}$
discovered in \cite{kr80}
\be
\ddot x_i=m^2\sum_{j\not=i}\wp'(x_i-x_j)
\leftrightarrow
\dot L(z)=[M(z),L(z)]
\ee
Here $L(z)$, $M(z)$ are $N\times N$ matrices given explicitly by
\bea
L_{ij}&=&\dot x_i\delta_{ij}-m(1-\delta_{ij})\Phi(x_i-x_j,z)\nonumber\\
M_{ij}&=&m\delta_{ij}\sum_{k\not=i}\wp(x_k-x_i)
+m(1-\delta_{ij})\Phi'(x_i-x_j,z)
\eea
It is not difficult to see that $L(z),M(z)$ is a Lax pair for the
$SU(N)$
Calogero-Moser system if the function $\Phi(x,z)$ satisfies the
following
elliptic functional equation
\be
\label{funeq}
\Phi(x,z)\Phi'(y,z)
-\Phi(y,z)\Phi'(x,z)
=
(\wp(x)-\wp(y))\Phi(x+y,z)
\ee
This functional equation is solved by
\be
\Phi(x,z)={\sigma(z-x)\over\sigma(z)\sigma(x)}\,
e^{x\zeta(z)}
\ee
where $\sigma(z)$, $\zeta(z)$ are the usual elliptic Weierstrass
functions. The fibration of spectral curves associated with the
Lax pair $L(z),M(z)$ can now be defined by
\be
\Gamma=\{(k,z);\ det\big(kI-L(z)\big)=0\},\ \ \
d\lambda=kdz
\ee
This fibration is the same as the fibration associated to the
$SU(N)$ Hitchin system \cite{martinecb, nekrasov}. In fact, the
spectral curve $\Gamma(\tau)$ is clearly invariant under the
transformation
$L(z)\to\tilde L(z)=GL(z)G^{-1}$, for $N\times N$ matrices $G$.
Choosing $G_{ij}=\delta_{ij}e^{x_i\zeta(z)}$,
we have $\Phi(x_i-x_j,z)\to
\tilde\Phi(x_i-x_j,z)
={\sigma(z-x_i+x_j)\over\sigma(z)\sigma(x_i-x_j)}=
-{1\over x_i-x_j}+\cdots$.
This leads to
$\tilde L_{ij}(z)=-m(1-\delta_{ij}){1\over z}+\cdots$, and
\be
\tilde L(z)-{m\over z}I
=-{m\over z}\pmatrix{1&\cdot \cdot& 1\cr
:& &:\cr
1&\cdot \cdot &1\cr}
\sim -{m\over z}\pmatrix{0&\cdot\cdot&0\cr
:& 0& :\cr
0& 0& N\cr}
\Rightarrow
\tilde L(z)\sim {m\over z}\mu+\cdots
\ee
This shows that $\tilde L(z)$ has exactly the same poles and
residues
as the Higgs field $\phi(z)$ of the $SU(N)$ Hitchin model.
The equivalence of the two fibrations follows.

\medskip
We check now the scaling limits. Clearly $m\to 0$ leads to a free
system, so we concentrate on the limit $m\to\infty$. This limit
was found in the late 1980's by Inozemtsev \cite{inozemtsev}, who
showed that if $m$ tends to $\infty$ according to the following rule
\be
m=M\,q^{-{1\over  2N}},
\ \ \
q=e^{2\pi i\tau}\to 0,
\ \omega_1=-i\pi
\ee
and if new dynamical variables $(X_i,P_i)$ are defined by
\be
x_i=X_i-2\omega_2{i\over N},
\ \ \
p_i=P_i
\ee
then the Hamiltonian $H_{CM}^{SU(N)}$ tends to the Hamiltonian
for the Toda system associated to the affine Lie algebra
$SU(N)^{(1)}$
\be
H_{CM}^{SU(N)}\to
H_{Toda}^{SU(N)^{(1)}}
=
{1\over 2}\sum_{i=1}^NP_i^2
-
{1\over 2}M^2\sum_{i=1}^{N}e^{X_{i+1}-X_i},
\ \ \
X_{N+1}\equiv X_1.
\ee
The basic mechanism
of this scaling limit is the following asymptotics for the
Weierstrass
$\wp$-function
\be
\wp(u)
={\eta_1\over i\pi}+\sum_{n=-\infty}^{\infty}
{1\over sh(u-2n\omega_2)-1}\sim
e^u+e^{-2\omega_2-u}
\ee
for $-2\omega_2<u<0$. In the potential
for the $SU(N)$ Calogero-Moser system, we may
assume without loss of generality that $i>j$.
Then $x_i-x_j=(X_i-X_j)-{2\omega_2\over N}(i-j)$
is in the range $(-2\omega_2,0)$, and we have
\be
m^2\wp(x_i-x_j)
=
M^2\big\{e^{X_i-X_j}e^{{2\omega_2\over N}(1-i+j)}
+
e^{-(X_i-X_j)}e^{{2\omega_2\over N}(1+i-j-N)}
\big\}
\ee
The only surviving terms arise from $x_{i+1}-x_i$, $1\leq i\leq
N-1$,
and $x_1-x_N$. This establishes the scaling limit of the
Hamiltonian
asserted above.
The Lax pair also admits a scaling limit
\be
L(z)\to L_{Toda}(Z), \ \ \ e^z=Zq^{-{1\over 2}},
\ \ \ q\to 0
\ee
since the function $\Phi(u,z)$ scales as
\be
\Phi(u,z)\to\cases{e^{-{1\over 2}u}(1-Z^{-1}e^{u-\omega_2}),
& if $\Re\,u\to +\infty$;\cr
-e^{{1\over 2}u}(1-Z^{-1}e^{-u-\omega_2}),
& if $\Re\,u\to -\infty$.\cr}
\ee
The existence of the scaling limit of the Lax pair insures the
existence of the scaling limit of the fibration of spectral
curves. In fact, it is easy to derive from the scaling limit
of $\Phi(u,z)$ an explicit formula for $L_{Toda}(Z)$, and
hence for the spectral curves of the Toda system
\be
\Gamma_{Toda}
=
\{(k,Z);\ {\rm det}(kI-L_{Toda}(Z))=0\}
\ee
We find ${\rm det}(kI-L_{Toda}(Z))=Z+{M^N\over Z}-P(k)$,
which is the Seiberg-Witten curve for the pure Yang-Mills theory.
Thus the consistency of the scaling limit as $m\to\infty$ has
also been verified from the point of view of integrable models. 

We have now recaptured the essential features of the Donagi-Witten
solution of the $SU(N)$ gauge theory in terms of Calogero-Moser
systems. It turns out that the Calogero-Mose system formulation also
provides another important check, namely that the effective
prepotential ${\cal F}$ defined by its spectral curves does
satisfy the logarithmic singularities required of the effective
prepotential for the $SU(N)$ gauge theory. This is because the
$SU(N)$ Calogero-Moser spectral curves turn out to admit a
parametrization in closed form, from which the classical
order parameters of the four-dimensional gauge theory
can be read off, and the perturbative expansion of the
periods $a_{Dk}$ evaluated \cite{dp98d}.

\subsection{$G$ Calogero-Moser systems and $G^{(1)}$ affine Toda systems}

The $SU(N)$ Calogero-Moser system admits a generalization to any
simple Lie group $G$, as introduced by Olshanetsky and Perelomov
\cite{olspe75} in the mid 1970"s
\be
H_{CM}^G
={1\over 2}\sum_{i=1}^{n}p_i^2
-{1\over 2}\sum_{\alpha\in {\cal
R}(G)}m_{|\alpha|}^2\wp(\alpha\cdot x)
\ee
where $m_{|\alpha|}$ is a mass parameter which depends only on
the length of the root $\alpha$. It is natural to look to these
systems
for the solution of the supersymmetric four-dimensional gauge
theory
with gauge group $G$, but for this, we need to examine their
scaling
limits. For $SU(N)$, we have seen that
the root lattice of $SU(N)$ reduces to the set of simple
roots for the affine Lie algebra $SU(N)^{(1)}$.
This scaling limit can be generalized to all simple Lie algebras
as follows \cite{dp98b}. We have
\be
\label{affinetoda}
m^2\sum_{\alpha\in {\cal R}(G)}\wp(\alpha\cdot x)
\to M^2\sum_{\alpha\in {\cal R}(G^{(1)})\atop
\alpha\ {\rm simple}}e^{\alpha\cdot X}
\ee
if $m^2$, $x_i$, and $p_i$ scale according to
\bea
\label{scaling}
m^2&=&M^2 q^{-{1\over h_G}}\nonumber\\
x&=& X-{2\omega_2\over h_G}\rho^{\vee},
\ \ p=P
\eea
where $\rho^{\vee}$ is the level vector and $h_G$ is the
Coxeter number of $G$. The right hand side of (\ref{affinetoda}) defines the
Hamiltonian
of the Toda system associated to the affine Lie algebra $G^{(1)}$,
so
that our result can be restated simply as
\be
H_{CM}^G\to H_{Toda}^{G^{(1)}}
\ee

\subsection{$G$ Twisted Calogero-Moser systems and $(G^{(1)})^{\vee}$
affine Toda systems}

The scaling limit described above for the $G$ Calogero-Moser
system
is however not suitable for the Seiberg-Witten solution of the
supersymmetric gauge theory with gauge group $G$.
According to the renormalization group, the decoupling of
the matter hypermultiplet should rather obey the following
scaling law
\be
\label{scalingv}
m^2=M^2 q^{-{1\over h_G^{\vee}}}
\ee
where $h_G^{\vee}$ is the dual Coxeter number of $G$.
When $G$ is not simply-laced, we have $h_G^{\vee}<h_G$, and the
$G$ Calogero-Moser system does not admit a finite limit
under this scaling. Thus new generalizations of the
$SU(N)$ elliptic Calogero-Moser system admitting finite
limits under the scaling (\ref{scalingv}) are required.
It turns out that these new systems are the {\it twisted
$G$ Calogero-Moser systems} defined as follows \cite{dp98a}
\be
H^{G}_{twisted}
=
{1\over 2}P^2-{1\over 2}\sum_{\alpha\in {\cal R}(G)}
m_{|\alpha|}^2\wp_{\nu(\alpha)}(\alpha\cdot x)
\ee
where $\nu(\alpha)=1$ if $\alpha$ is a long root,
$\nu(\alpha)=2$ if $\alpha$ is a short root of $B_n$, $C_n$,
$F_4$,
and $\nu(\alpha)=3$ if $\alpha$ is a short root of $G_2$.
The key to the twisting is the improved asymptotics
for the twisted Weierstrass $\wp$-function
\be
\wp_{\nu}(u)
=
\sum_{k=0}^{\nu-1}\wp(u+2\omega_2{k\over\nu})
=
{\nu^2\over 2}\sum_{n=-\infty}^{\infty}{1\over {\rm ch}\,\nu(u-2n\omega_2)-1},
\ \ \nu=1,2,3.
\ee
Then under (\ref{scalingv}), with new dynamical variables
$X_i,P_i$
defined by
\be
x=X-{2\omega_2\over h_G^{\vee}}\rho,
\ \ p=P
\ee
where $\rho$ is the Weyl vector,
the improved asymptotics lead to the finite limit
\be
H^G_{twisted}\to
H_{Toda}^{(G^{(1)})^{\vee}}
\ee
where $H_{Toda}^{(G^{(1)})^{\vee}}$ is the Toda Hamiltonian
associated to the dual of the affine Lie algebra $G^{(1)}$.
The emergence of $H_{Toda}^{(G^{(1)})^{\vee}}$ as the limit
of the twisted $G$ Calogero-Moser system is another confirmation
of
the latter system as the solution of the $G$ gauge theory with
an adjoint hypermultiplet. Indeed, generalizing the case of $SU(N)$, the system $H_{Toda}^{(G^{(1)})^{\vee}}$
has been shown by Martinec and Warner \cite{martinec} to be the
solution
of the supersymmetric pure Yang-Mills theory. Conversely, this
result
would follow from the solution of the theory with adjoint
hypermultiplet by twisted Calogero-Moser systems.

\subsection{Lax pairs with spectral parameter for Calogero-Moser systems}

The twisted $G$ Calogero-Moser systems have now been shown to be
the
correct models for the Seiberg-Witten solution of the $G$ gauge
theory with an adjoint hypermultiplet. However, for all Lie groups
except $SU(N)$, we still have to
establish their integrability and the existence of a Lax pair with
spectral parameter. Even the integrability of the untwisted $G$
Calogero-Moser systems had not fully been established prior to
\cite{dp98a}.
The only known results at that time went back to the 1975 work of
Olshanetsky and Perelomov \cite{olspe75}, and were as follows

$\bullet$ $G$ classical ($B_n,C_n$, or $D_n$): a Lax pair was
known, but without
spectral parameter.

$\bullet$ $G$ exceptional ($G_2,F_4,E_6,E_7,E_8$): no Lax pair was
known.

\noindent
The integrability of twisted Calogero-Moser systems had of course
not
even been an issue at this point. It turned out that all these
systems
admit Lax pairs with spectral parameter, and we now describe these
Lax pairs.

\subsubsection{Construction of the Lax pairs}

A major difficulty in the search for a Lax pair in the case of a general simple Lie algebra $G$ is that, unlike in
the case of $SU(N)$, it
cannot be found in the Lie algebra $G$.
Rather, we proceed as follows.

\medskip
Let $\Lambda:\ G \longrightarrow \ GL(N,{\bf C})$ be a
representation of $G$ of dimension $N$,
$\lambda_I$ ($I=1,\cdots,N$) its weights, and
\be
\alpha_{IJ}=\lambda_I-\lambda_J\
\ee
the weights of $\Lambda\otimes \Lambda^*$.
Let $h_i$, $i=1,\cdots,n$ be generators of the Cartan subalgebra
${\cal H}_G$ of $G$, and let $\tilde h_j$, $j=n+1,\cdots,N$,
satisfy
$[h_i,\tilde h_j]=[\tilde h_i,\tilde h_j]=0$. Let ${\cal H}$ be
generated by $h_i\oplus \tilde h_j$. Let $u_I$ be the weights of
the fundamental representation of $GL(N,{\bf C})$. We can write
\be
su_I=\lambda_I+v_I\ \ {\rm with}\ \ \lambda_I\perp v_J,
\ee
where $s^2={1\over n}\sum_{I=1}^N\lambda_I\cdot\lambda_I$ is the
Dynkin index. Let $E_{IJ}=u_Iu_J^T$ be the generators of
$GL(N,{\bf C})$.
We now look for a Lax pair $L(z),M(z)$ under the following form
\be
L=P+X,\ \ \ M=D+Y
\ee
where
\bea
X&=&\sum_{I\not=J}C_{IJ}\Phi_{IJ}(\alpha_{IJ}\cdot x,z)E_{IJ}
\nonumber\\
P&=& p\cdot h\nonumber\\
Y&=&\sum_{I\not=J}C_{IJ}\Phi_{IJ}'(\alpha_{IJ}\cdot x,z)E_{IJ}
\nonumber\\
D&=&d(h\oplus \tilde h)+\Delta
\eea
with the coefficients $C_{IJ}$ and the functions $\Phi_{IJ}(x,z)$
yet
to be determined. We observe that the case of $SU(N)$
corresponds to all functions
$\Phi_{IJ}(u,z)$ equal to $\Phi(u,z)$, and that the coefficients
$C_{IJ}$
are equivalent to the matrix of residues $\mu$. In the general
case,
they have to be solved for. To do so, we introduce the following
notation
\bea
\Phi_{IJ}&=&\Phi_{IJ}(\alpha_{IJ}\cdot x,z)\nonumber\\
\wp_{IJ}'&=&\Phi_{IJ}(\alpha\cdot x,z)\Phi_{JI}'(-\alpha\cdot x,z)
-
\Phi_{IJ}(-\alpha\cdot x,z)\Phi_{JI}'(\alpha\cdot x,z)
\eea
Then the matrices $L(z)$, $M(z)$ are a Lax pair for the (twisted
or
untwisted) Calogero-Moser system if and only if the following
functional
equations are satisfied
\bea
\sum_{I\not=J}C_{IJ}C_{JI}\wp_{IJ}'\alpha_{IJ}
&=&
s^2\sum_{\alpha\in {\cal
R}(G)}m_{|\alpha|}^2\wp_{\nu(\alpha)}(\alpha\cdot x)
\nonumber\\
\sum_{I\not=J}C_{IJ}C_{JI}\wp_{IJ}'(v_I-v_J)
&=&
0
\nonumber\\
\sum_{K\not=I,J}C_{IK}C_{KJ}(\Phi_{IK}\Phi_{KJ}'-
\Phi_{IK}'\Phi_{KJ})
&=&
\sum_{K\not=I,J}\Delta_{IJ}C_{KJ}\Phi_{KJ}
-
\sum_{K\not=I,J}C_{IK}\Phi_{JK}\Delta_{KJ}
\nonumber\\
&&
\ \ \ \ +sC_{IJ}\Phi_{IJ}d\cdot (v_I-v_J)
\eea
More specifically, it turns out that the condition $\dot x=p$ is
equivalent to $\dot X=[P,Y]$. The second condition above combined
with
the Calogero-Moser equations of motion are equivalent to
$\dot P=[X,Y]_{\cal H}$, where the subscript ${\cal H}$ denotes
projection onto that subspace. Finally, the third condition above
is equivalent to $[X,Y]_{GL(N,{\bf C})\ominus {\cal H}}
+[X,d\cdot(h\oplus\tilde h)+\Delta]=0$.

\bigskip
\noindent
{\bf Theorem 1.} \cite{dp98a}. {\it Lax pairs $L(z),M(z)$ with
spectral
parameter $z\in {\bf C}/{\bf Z}+\tau{\bf Z}$ of the above form can
be found for both twisted and untwisted Calogero-Moser systems,
for all simple Lie algebras $G$, except possibly in the case of
twisted $G_2$.
In the case of $E_8$, we have to assume the existence of a sign
assignment satisfying a cocycle-type condition.}

\subsubsection{The scaling limit of the Lax pair}

The basic property of the Lax pairs for the twisted and untwisted
Calogero-Moser systems which we just constructed is

\bigskip
\noindent
{\bf Theorem 2.} \cite{dp98b}. {\it All Lax pairs constructed above
admit
a finite scaing limit. More precisely, the limit is taken with
respect to the scaling law (\ref{scaling}) for untwisted
Calogero-Moser systems, and with respect to \ref{scalingv} for
twisted Calogero-Moser systems.}

\bigskip
We have seen that the scaling limit of the Calogero-Moser
Hamiltonian
was a consequence of some precise asymptotics for the Weierstrass
$\wp$-function. For the Lax pairs, we need the precise asymptotics
of the
functions $\Phi_{IJ}(u,z)$.
In the case of untwisted Calogero-Moser systems, set
\be
C_{IJ}=M_{|\alpha|}e^{\delta\omega_2}c_{IJ},
\ \ \
Z=e^zq^{-{1\over 2}}
\ee
All the functions $\Phi_{IJ}(u,z)$ are given in this case by
$\Phi(u,z)$, and the scaling limit of the Lax pair $L(z)$, $M(z)$
followed from
\bea
L&:& C_{IJ}\Phi(\alpha\cdot x,z)
\to \cases{\pm c_{IJ} e^{\mp {1\over 2}\alpha\cdot X},
& if $l(\alpha)=\pm 1$;\cr
\mp c_{IJ} e^{\pm {1\over 2}\alpha\cdot X},
& if $l(\alpha)=\pm l_0$;\cr
0 & if otherwise.\cr}
\\
M&:& C_{IJ}\Phi'(\alpha\cdot x,z)
\to {1\over 2}\epsilon_{\alpha}\,{\rm lim}\,
C_{IJ}\Phi(\alpha\cdot x,z),
\ \
\epsilon_{\alpha}
=
\cases{1, & if $l(\alpha)=l_0$ or $l(\alpha)=-1$;\cr
-1, & if $l(\alpha)=-l_0$ or $l(\alpha)=1$.\cr}
\nonumber
\eea
In the case of twisted Calogero-Moser systems, set instead
$C_{IJ}=M_{|\alpha|}e^{\delta^{\vee}\omega_2}c_{IJ}$.
In this case, several distinct functions $\Phi_{IJ}(u,z)$ arise
\bea
&&\Phi_1(u,z)=\Phi(u,z)-\Phi(u+\omega_1,z)e^{\pi
i\zeta(z)+z\zeta(\omega_1)}
\nonumber\\
&&\Phi_2(u,z)={\Phi(u,z)\Phi(u+\omega_1,z)\over \Phi(\omega_1,z)}
\eea
as well as $\Phi_2(u+\epsilon_{IJ}\omega_2,z)$
where $\epsilon_{IJ}=\pm1$.
The existence of the scaling limit of the Lax pairs for
the twisted Calogero-Moser systems follows from the scaling limits
of $\Phi_1(u,z)$, $\Phi_2(u,z)$,
\bea
\Phi_1(u,z) &\to & \mp 2Z^{\mp 1}e^{\pm {1\over 2}u-\omega_2},
\ \ u\to \pm\infty
\nonumber\\
\Phi_2(u,z) &\to &
\pm 2 e^{\mp u}(1-Z^{\mp 1}e^{\pm 2u-2\omega_2}),
\ \ u\to\pm\infty
\eea

\subsection{Exact solution for super Yang-Mills with adjoint
hypermultiplet}

The Lax pair $L(z),M(z)$ of the twisted $G$ Calogero-Moser system
provides now the Seiberg-Witten solution of the ${\cal N}=2$
supersymmetric gauge theory with gauge group $G$ and a
hypermultiplet
in the adjoint representation \cite{dp98c}. It is given by
\be
\Gamma=\{(k,z);\ det\big(kI-L(z)\big)=0\},\ \ \
d\lambda=kdz
\ee
This can be checked explicitly for low $D_n$ and low rank $n$,
by working out the logarithmic singularities of the above
fibration
in the trigonometric limit. In this limit,
\be
\wp(z)\to {1\over Z^2}-{1\over 6},
\ \
\Phi(x,z)\to {1\over 2}{\rm coth}\,{1\over 2}x-{1\over Z},
\ \
{1\over Z}={1\over 2}{\rm coth}\,{1\over 2}x
\ee
and the equation for the fibration simplifies considerably
\be
{\rm det}\,(kI-L(z))
=
{m^2+mA-2k{m\over z}\over m^2+2mA}H(A)
+
{mA+2k{m\over Z}\over m^2+2mA}H(A+m)
\ee
Here the moduli vacua are parametrized now by monic polynomials
$H(A)$ of the form $H(A)=\prod_{j=1}^n(A^2-p_j^2)$,
and the variable $A$ is related to the variables $(k,z)$ by
\be
A^2+mA+2k{m\over Z}-k^2=0
\ee
The Seiberg-Witten form $d\lambda$ can then be re-expressed as
\be
d\lambda
=
-Adu,
\ \ \
e^u={(k+A+m)(k-A-m)\over k^2-A^2}
\ee
The methods of \cite{dp98d} lead then to the desired prepotential
\be
{\cal F}
=-{1\over 8\pi i}
\sum_{\alpha\in{\cal R}(D_n)}
(\alpha\cdot\phi)^2{\rm ln}\,(\alpha\cdot\phi)^2
-
(\alpha\cdot\phi+m)^2{\rm ln}\,(\alpha\cdot\phi+m)^2
\ee

\subsection{Other developments}

There has been many related developments since the works
\cite{dp98a, dp98b, dp98c}.

Seiberg-Witten solutions for the ${\cal N}=2$ gauge theory with
gauge group $G$ and several hypermultiplets in the adjoint
representation have been proposed by Uranga and Yokono
\cite{urayo}, under the assumption
of total zero mass.

Lax pairs with spectral parameter for both twisted and untwisted
Calogero-Moser systems, including the case of twisted $G_2$, have
been
constructed by Bordner, Sasaki, Corrigan, et al. \cite{bordner}.
These Lax pairs are different from the ones constructed in
\cite{dp98a}.
In particular, they do not admit finite scaling limits under
(\ref{scalingv}), and are not candidates for the Seiberg-Witten
solution
of the $G$ gauge theory with adjoint hypermultiplet.

It is intriguing that, except for $SU(N)$, the Lax pairs obtained
so far do not fit in the framework of classical Hitchin systems.
Recently, Hurtubise and Markman \cite{hurtubi} have proposed a new
general for modified Hitchin systems, which can incorporate both
Lax pairs constructed in \cite{dp98a} and in \cite{bordner}, at least for
the case of untwisted Calogero-Moser systems.

The Calogero-Moser systems can be viewed as non-relativistic
limits
of Ruijsenaars-Schneider systems \cite{ruij}. Except for $SU(N)$, the
integrability of these systems is still incompletely understood.
It may be hoped that advances in the theory of Calogero-Moser
systems
may help advances on Ruijsenaars-Schneider systems. Recently,
there
has been progress on the integrability of Ruijsenaars-Schneider
systems
for certain classical algebras, thanks to the works of B.Y. Hou et
al.
\cite{hou}.

Ruijsenaars-Schneider systems have been proposed by H. Braden,
A. Marshakov, A. Mironov, and A. Morozov \cite{marshakov} as the
Seiberg-Witten
solution of supersymmetric gauge theories in dimensions $5$ and
$6$,
with the relativization coming from the contributions of the
infinite
tower of Kaluza-Klein modes. This is an imprtant direction of
investigation, since supersymmetric gauge theories in $5$ and $6$
dimensions are difficult to explore otherwise.

Relations with the E-string and the reduction  on $T^2$ of
the 6-dimensional (2,0) theory may be found in \cite{eguchi}.
Recently, Dijkgraaf and Vafa
\cite{dijkgraaf} have proposed a promising relation between
the effective prepotential for supersymmetric gauge theories
and matrix models. Some developments in this direction and
related to the issues discussed here can be found in
\cite{matrix}. 

\section{New Spin Chain Models from M Theory}
\setcounter{equation}{0}

In the previous section, we have seen how integrable models can
produce the Seiberg-Witten exact solution of a supersymmetric
gauge theory.
The correspondence can also go the other way: here we
discuss
how the Seiberg-Witten solution of a gauge theory, in this case
the $SU(N)$ theory with a hypermultiplet in either the symmetric
or the
antisymmetric representation, can lead to new integrable models.

\subsection{A periodic generalized spin chain model}

We consider first the case of a hypermultiplet in the
antisymmetric
representation. Recall that the fibration $\Gamma(\Lambda)$ and
differential $d\lambda$ have been found by Landsteiner and Lopez,
and are given by (\ref{LL}). Postponing for the moment the choice of
$d\lambda$, our problem is to find $L(x)$, $M(x)$
satisfying a Lax equation, with the spectrum of $L(x)$ determined
by $\Gamma(\Lambda)$. We shall look for such a Lax pair in a
(generalized) spin chain model \cite{kp00}.
Henceforth, we set $\Lambda=1$ for notational simplicity.

\medskip
In general, an integrable chain model $\{\psi_n\}_{n\in{\bf Z}}$
is defined as the compatibility
condition for two linear equations of the form
\be
\lbrace\matrix{\psi_{n+1}=L_n(x)\psi_n\cr
\dot\psi_n=M_n(x)\psi_n\cr}\rbrace
\ \Rightarrow \
\dot L_n(x)=
M_{n+1}(x)L_n(x)-L_n(x)M_n(x)
\ee
Under the periodicity condition $L_{N+2}(x)=L_0(x)$,
$M_{N+2}(x)=M_0(x)$, the equation on the right hand side implies
the Lax pair equation
\be
\dot L(x)=[M(x),L(x)]
\ee
where $L(x)$ is defined by $L(x)=L_{N+1}(x)L_N(x)\cdots L_0(x)
\equiv\prod_{j=0}^{N+1}L_j(x)$,
and $M(x)=M_0(x)$. The spectral curve $\Gamma$ of such an
integrable system can then be defined as usual by
\be
\Gamma=\{(x,y);\ {\rm det}\,(yI-L(x))=0\}
\ee
Returning to the construction of the desired integrable model,
the key property is the invariance of the Landsteiner-Lopez curve
under
the involution
\be
\sigma\ :\ (x,y)\ \leftrightarrow \ (-x,y^{-1})
\ee
We see then that the spectral curve $\Gamma$ would reproduce the
Landsteiner-Lopez curve if the matrix $L(x)$ is $3\times 3$, and
satisfies
\be
{\rm det}\,L(x)=1,
\ \
L(x)^{-1}=-L(x),
\ \
Tr\,L(x)=3+{\cal O}(x^2)
\ee
In analogy with the $2\times 2$ Lax matrix used in \cite{korchem}
for the integration of a quasi-classical approximation to a
system of reggeons in $QCD$, we can achieve this by setting
\be
L_n(x)=1+x\,s_ns_n^T,
\ \
M_n(x)=x{1\over s_n^Ts_{n+1}}(s_{n-1}s_n^T+s_ns_{n-1}^T)
\ee
where $s_n$ is a periodic sequence of complex $3$-vectors
satisfying $s_n^Ts_n=0$, $s_{n+N+2}=s_n$. We obtain in this way
an integrable model, with $s_n$ as dynamical variables satisfying
the equation of motion
\be
\dot s_n={s_{n+1}\over s_{n+1}^Ts_n}
-
{s_{n-1}\over s_{n-1}^Ts_n}
\ee
This integrable model admits the same spectral curves as the
$SU(N)$
gauge theory with a hypermultiplet in the antisymmetric
representation.
However, it turns out that the associated differentials
$d\lambda$ in the two theories do not coincide. The reason is the
parity
of the differential $d\lambda$ under the involution
$(x,y)\rightarrow
(-x,y^{-1})$, which requires instead the following model
\be
\label{pnqna}
\dot p_n={p_{n+1}\over p_{n+1}^Tq_n}+{p_{n-1}\over
p_{n-1}^Tq_n}+\mu_np_n,
\ \
\dot q_n=-{q_{n+1}\over p_n^Tq_{n+1}}-{q_{n-1}\over p_n^Tq_{n-1}}
-\mu_nq_n
\ee
Here $\mu_n$ is an arbitrary multiplier,
and the dynamical variables $q_n,p_n$ are complex $3$-vectors,
satisfying the following conditions
\bea
&&q_{n+N+2}=q_n,\ \ p_{n+N+2}=p_n\\
&&p_n^Tq_n=0,\ \ p_n=g_0p_{-n-1},
\ \ q_n=g_0q_{-n-1}
\eea
where $g_0$ is the $3\times 3$ diagonal matrix with
$g_{ii}=(-1)^{i+1}$. Then the system admits a Lax pair $L(x)$,
$M(x)$
given by
\be
L(x)=\prod_{n=0}^{N+1}(1+x q_np_n^T),\ \ \
M(x)=x\left({q_{N+1}p_0^T\over p_0^Tq_{N+1}}-{q_0p_{N+1}^T\over
p_{N+1}^Tq_0}\right)
\ee
The corresponding spectral curve $\Gamma =\{(x,y);{\rm
det}\,(yI-L(x))=0\}$ is the Landsteiner-Lopez curve. Furthermore,
there is a correspondence between the variables $(q_n,p_n)$ and
pairs $(\Gamma,[D])$
\be
\label{phase}
(q_n,p_n)\ \leftrightarrow\ (\Gamma,[D])
\ee
where $[D]=[z_1,\cdots,z_{2N+1}]$ is a divisor even under the
involution $\sigma$. For given $(q_n,p_n)$, $[D]$ is the divisor
of
poles of the Bloch eigenfunction
$\psi_n(x,y)$ defined by $\psi_{n+1}(x,y)=L_n(x)\psi_n(x)$,
$\psi_{n+N+2}(x,y)=y\psi_n(x,y)$, for $(x,y)\in\Gamma$.
We shall denote also denote the points $(x,y)$ on $\Gamma$ by $Q$.
Let the action variables $a_i$ and the angle variables $\phi_i$
be defined on the $2(N-1)$-dimensional space $\M_0$ by
\be
a_i=\oint_{A_i}d\lambda,
\quad\quad
\phi_i=\sum_{i=1}^{2N+1}\int^{z_i}d\omega_i
\ee
where $\{A_i\}_{1\leq i\leq N-1}$ and $\{d\omega_i\}_{1\leq i\leq
N-1}$, are respectively a basis for the even cycles and a basis
for the even holomorphic differentials on $\Gamma$. Then
the form $\omega=\delta(\sum d\lambda(z_i))$
defines a symplectic form on $\M_0$ which can also be expressed as
\be
\omega=\sum_{i=1}^{N-1}\delta a_i\wedge\delta\phi_i\label{sf0}
\ee
The dynamical system (\ref{pnqna}) is Hamiltonian with respect to
this symplectic form, with Hamiltonian
\be
H=u_{N-2}=\sum_{n=0}^{N+1}
{(p_n^Tq_{n-3})\over
(p_n^Tq_{n-1})(p_{n-1}^Tq_{n-2})(p_{n-2}^Tq_{n-3})}-
{(p_n^Tq_{n-2})^2\over 2(p_n^Tq_{n-1})^2(p_{n-1}^Tq_{n-2})^2}
\ee
Thus the system (\ref{pnqna}) is the integrable model that we were
looking for. The correspondence between the Hamiltonian structure 
for
the dynamical variables $(q_n,p_n)$ and the geometric symplectic
form $\omega$ for $(\Gamma,[D])$
depends fundamentally on the fact that they both
coincide with a symplectic form
which can be defined in terms of the Lax pair
\be
\label{oasym}
\omega
=
{1\over 2}\sum_{\alpha=1}^3
{\rm Res}_{P_{\alpha}}\langle \Psi_{n+1}^*(Q)\delta L_n(x)\wedge
\delta\Psi_n(Q)\rangle dx
\ee
where $P_{\alpha}$ are the points in $\Gamma$ lying above
$x=\infty$.

\subsection{Models with twisted monodromies}

We turn now to the problem of finding an integrable model
corresponding to the Seiberg-Witten solution of the $SU(N)$ gauge
theory with matter
in the symmetric representation. In this case, the spectral curves
still admit an involution $\sigma$, but which is now of the form
\be
\sigma\ :\ (x,y)\ \longrightarrow \ (-x,x^4y^{-1})
\ee
As suggested earlier \cite{kp97}, such shifts are indicative of
twisted monodromy conditions. We consider then the following
dynamical
system \cite{kp02}
\bea
\label{dynsytw}
\dot p_n&=&
{p_{n+1}\over p_{n+1}^Tq_n}+{p_{n-1}\over p_{n-1}^Tq_n}+\mu_np_n,
\quad
\dot q_n=
-{q_{n+1}\over p_n^Tq_{n+1}}-{q_{n-1}\over p_n^Tq_{n-1}}-\mu_np_n
\nonumber\\
\dot a&=&
\{{q_{m-1}p_m^T
\over p_m^Tq_{m-1}}-{q_mp_{m-1}^T
\over p_{m-1}^Tq_m},b\},\quad\quad
\dot b=\{{q_{m-1}p_m^T
\over p_m^Tq_{m-1}}-{q_mp_{m-1}^T
\over p_{m-1}^Tq_m},c\},\quad\quad
\dot c=0
\eea
Here $\mu_n(t)$ is again an arbitrary scalar function, and we have
set
$m=-{N\over 2}+1$ for $N$ even and $m=-{N\over 2}+{1\over 2}$
for $N$ odd. The variables $q_n,p_n$, $a$, $b$, $c$ are all
$3\times 3$
matrices satisfying
\bea
&&q_n^Tp_n=0,\ \ p_n=hp_{-n-1},\ \ q_n=hq_{-n-1},
\\
&&
a^2=1,\ \ ab=ba,\ \ b^2=ac+ca,\ \ bc=cb,\ \ c^2=0
\eea
where $h$ is the $3\times 3$ matrix whose only non-zero entries
are $h_{31}=h_{22}=h_{13}=1$.
The above system appears uncoupled, but it will
not be after imposing twisted monodromy conditions
on $(q_n,p_n)$. More precisely,
let $L_n(x)=1+x q_np_n^T$ as before. Then

$\bullet$ There are unique $3\times 3$ matrices
$g_n(x)=a_nx^2+b_nx+c_n$ which satisfy the periodicity condition
\be
g_{n+1}L_{n+N-2}=L_ng_n
\ee
for any fixed data $a_r,b_r,c_r,
(p_n,q_n)_{n=r}^{n=r+N-3}$ with the constraint $q_n^Tp_n=0$.

$\bullet$ The above dynamical system
with $a_m=ah, b_m=bh, c_m=ch$ is
integrable, in the sense that it is equivalent to
the following Lax equation
\be
\dot L_n=M_{n+1}L_n-L_nM_n,
\ \ \
M_n(x)\equiv x\bigg({q_{n-1}p_n^T
\over p_n^Tq_{n-1}}-{q_np_{n-1}^T\over p_{n-1}^Tq_n}\bigg)
\ee

$\bullet$ The spectral curve $\Gamma=\{(x,y);det\big(yI-g_n(x)
L_{n+N-3}(x)\cdots L_n(x)\big)=0\}$
is independent of $n$ and
coincides with the
Landsteiner-Lopez curve (\ref{LLs}).
The dynamical system (\ref{dynsytw}) is Hamiltonian with
respect to the symplectic form $\omega=\sum_{i=1}^{N-2}\delta x(z_i)
\wedge {\delta y\over y}(z_i)$
on the reduced phase space $u_N=1$, $u_{N-1}=0$,
through the usual correspondence between dynamical variables
and curves and divisors.
The Hamiltonian is $H=u_{N-2}$.

$\bullet$ The symplectic form $\omega$ can also be expressed in
terms of the Lax operator $L_n(x)$ and the matrices $g_k(x)$
defining the twisted monodromy conditions
\be
\label{osym}
\omega=
{1\over 2}
\sum_{\alpha=1}^3
\res_{P_{\alpha}}
(\langle \psi_{n+1}^*(Q)\delta L_n(x)\wedge
\delta\psi_n(Q)\rangle_k
+
\psi_k^*(\delta g_k\,g_k^{-1})\wedge \delta\psi_k)\,dx
\ee
Here $\langle f_n\rangle_k$ is defined to be
$\sum_{n=k}^{N+k-3}f_n$, $\psi_n(Q)$ is the Baker-Akhiezer function,
and $\psi_n^*(Q)$ the dual Baker-Akhiezer function characterized by
\be
\psi_{n+1}^*(Q)L_n(x)=\psi_n^*(Q),
\ \
\psi_{k+N-2}^*g_k^{-1}(Q)=y^{-1}\psi_n^*(Q),
\ \
\psi_k^*(Q)\psi_k(Q)=1.
\ee
In summary, we have

\medskip
\noindent
{\bf Theorem 3.} \cite{kp00, kp02} {\it The Seiberg-Witten solution
for both $SU(N)$ gauge theories with a hypermultiplet in either
the symmetric or the antisymmetric representation can be realized
as the spectral curves and symplectic form of a Hamiltonian system
admitting a Lax pair
representation. These systems are given by spin chains, with
twisted monodromy in the case of the symmetric representation.
The Hamiltonian for each model turn is
the beta function of the corresponding gauge theory}.

\bigskip

\section{Hamiltonian Formulation of Soliton Equations}
\setcounter{equation}{0}

The correspondence between ${\cal N}=2$ supersymmetric gauge
theories and integrable models begins with the identification of
their spectral curves. However, as we have seen in the case of the
$SU(N)$ with a hypermultiplet in the antisymmetric representation,
the identification of spectral curves has to be supplemented by an
identification of symplectic structures. More precisely, on the
gauge theory side, the Seiberg-Witten meromorphic form $d\lambda$
equips a moduli space of spectral curves and divisors with a
symplectic form. On the integrable model side, the phase space
must be then equipped with a corresponding symplectic structure
with respect to which the integrable model is Hamiltonian. What is
this symplectic structure? Since the integrable model is given by
its Lax pair, what is needed is a general construction of a
Hamiltonian structure for an integrable model directly from its
Lax representation. Such a construction was found in \cite{kp97}
\cite{kp98}, and developed further in \cite{kr99}. The essential
features of this symplectic form are summarized in the following
general formula
\be
\label{symplfo}
\omega
=
{1\over 2}\sum_{\alpha}\res_{\alpha}\langle \psi^{\dagger}(x,k)\delta L(x)
\wedge \delta\psi(x,k)\rangle dk
\ee
The set-up for this formula is broadly as follows.
The phase space of the system is a space ${\cal L}$ of operators $L(x)$.
The operators $L(x)$ can be finite-dimensional matrices, or
differential operators in the variable $x$.
The expression $\psi(x,k)$ is the
Baker-Akhiezer (or Bloch) function, which is 
an eigenvector of $L(x)$. The variable $k$ is the
spectral parameter, or an analogous quantity
\footnote{The variable $k$ here is the analogue of the variable
$x$ of \S 4, and of the variable $z$ of \S 3.
The notation of (\ref{symplfo}) is consistent with the case of the
Korteweg-deVries equation, where $x$ corresponds
to the variable in $L(x)=\p_x^2+u(x)$.
Unfortunately, the diversity of integrable models and entrenched practices makes
attempts at a uniform notation impractical.}.
The expression $\psi^{\dagger}(x,k)$ is the analogous dual Baker-Akhiezer
function, which should be viewed as a row vector, if $\psi$
is a column vector. The functions $\psi(x,k)$,
$\psi^{\dagger}(x,k)$ have compensating essential singularities in $k$,
so that all combined expressions are
meromorphic. 
The differential $\delta$ denotes exterior
differentiation on the space ${\cal L}$,
and $\langle\cdot\rangle$ denotes averaging with respect to all
variables $x$. The points $P_{\alpha}$ are given fixed poles.
Thus (\ref{symplfo}) defines a 2-form on the phase space ${\cal L}$.

\medskip
The above formula provides a universal framework for
the construction of symplectic forms in terms of the Lax operator
$L(x)$. In practice, for each integrable model, a suitable ambiant
space ${\cal L}$ for the operators $L(x)$ has to be specified,
in order for all terms in (\ref{symplfo}) to have the desirable
meromorphicity properties and the resulting formula to be independent of the normalization for the Baker-Akhiezer functions. 
The integrable model is then obtained by
constructing an operator-valued function $M(x)$ on ${\cal L}$
and considering the corresponding flow
\be
\p_tL(x)=[L(x),M(x)]
\ee
which will be Hamiltonian with respect to $\omega$. This was done
in some generality in \cite{kp97, kp98}, where ${\cal L}$ was
constructed from the leaves in a foliated moduli space of curves
with two specified Abelian integrals, as well as from spaces of
ordinary differential operators. It is also evident that the
symplectic forms found in \S 4 in the context of the $SU(N)$ gauge
theories with symmetric or antisymmetric hypermultiplets are examples
of the same construction. It may be helpful to have an intermediate
framework which is at the same time sufficiently specific for easy use,
and broad enough to encompass all models encountered in supersymmetric
gauge theories. We present such a framework below, together with
a brief description of some further recent developments.

\subsection{The universal symplectic form}

For finite-dimensional integrable systems which admit a Lax pair representation with a rational spectral parameter, a practical
framework is the following.

\medskip

Let $\L(D)$ be a space of meromorphic matrix functions
\begin{equation}
L(z)=u_0+\sum_{m=1}^n\sum_{l=1}^{h_m}{u_{ml}\over(z-z_m)^{l}}
\end{equation}
with a fixed divisor of poles $D=\sum_{m=1}^nh_mz_m$.
The integrable models we consider are 
flows on the space $\L(D)$ which can be constructed as follows.
For each $L\in \L(D)$, the matrix functions
$M_{n,p}(z,L)$ are defined by the formula
\beq
M_{n,p}(z,L)={L^n(p)\over z-p},\ \ p\neq z_m.
\eeq
Then the commutator $[M,L]$ has no pole at $z=p$.
If we identify the vector space $\L(D)$ with its own tangent space,
$[M,L]$ can be regarded
as a tangent vector field $\p_{n,p}$ to $\L(D)$. The 
corresponding flow on $\L(D)$ 
\beq\label{Lax1}
{\partial\over\partial t_{n,p}}L=[M_{n,p},L]
\eeq
is a dynamical system which admits by construction a 
Lax representation. Here $t_{n,p}$ is the time of the flow
defined by $M_{n,p}(z,L)$. Standard arguments from 
the theory of solitons show that all these flows commute with 
each other. We stress that the construction
of the flows on $\L(D)$ does not depend on a Hamiltonian 
structure.

\medskip
Next, we define a two-form on $\L(D)$ by the formula
\begin{equation}\label{form1}
\omega={1\over 2} \sum_{a}{\rm Res}_{z_a}{\rm Tr}
\left(\Psi^{-1} \delta L(z)\wedge \delta \Psi(z)\right) \,{dz}
\end{equation}
The sum is taken over the set of all the poles of $L$ 
together with the pole of $dz$
at $z_0=\infty$, i.e., $z_a=\{z_0, z_1,\cdots,z_n\}$.
We shall assume for simplicity that the normalization point
$z_0$ does not coincide with any of the other punctures $z_m$.
The case when $z_0$ coincides with one of the punctures
can be treated with only slight technical modifications.
The various components of the above formula are as follows.
The entries of matrices $u_0,u_{ml}$ can be viewed as coordinates on $\L(D)$. If we denote the exterior differentiation on $\L(D)$ by
$\delta$, then $\delta L(z)$ can be regarded as a matrix valued one-form on $\L(D)$
\begin{equation}
\delta L(z)=\delta u_0+\sum_m\sum_{l=1}^{h_m}{\delta u_{ml}\over(z-z_m)^{l}}
\end{equation}
Let $\Psi(z)$ be the matrix whose columns are {\it normalized}
eigenvectors of $L(z)$, i.e.
\begin{equation}
L(z)\Psi(z)=\Psi(z)K(z),\ \ e_0\Psi=e_0
\end{equation}
where $K$ is a diagonal matrix $K^{ij}=k_i\delta^{ij}$, and $k_i$ are
the eigenvalues of $L(z)$. The co-vector $e_0$ defining the normalization of the
eigenvectors is $e_0=(1,1\ldots,1)$. The external differential 
$\delta\Psi$ of $\Psi$
can be viewed as a one-form on $\L(D)$, and the formula (\ref{form1})
defines a two-form on $\L(D)$.

\medskip
A change of normalization vector $e_0$ leads to a transformation
\begin{equation}
\Psi(z)\to \Psi(z)'=\Psi(z)h(z)
\end{equation}
where $h(z)$ is a diagonal matrix. Under such transformation $\omega$ gets changed to
\begin{eqnarray}
\omega'&=&\omega+{1\over 2} \sum_{a}{\rm Res}_{z_a}
{\rm Tr}\left(\Psi^{-1} \delta L(z)\Psi(z)\wedge \delta h h^{-1}\right)
{dz}\nonumber\\
&=&\omega+{1\over 2} \sum_a{\rm Res}_{z_a}{\rm Tr}
\left(\delta K\wedge \delta h h^{-1}\right){dz}
\end{eqnarray}
We fix now a set of diagonal matrices
$C=(C_0,C_m)$
\beq
\label{C}
C_{0}(z)=C_{0,0}+C_{0,1}z^{-1},\ \ C_m(z)=\sum_{l=1}^{h_m}C_{m,l}(z-z_m)^{-l},\ m>0
\eeq
and define a subspace $\M=\M^C$ of $\L(D)$ by the constraints
\begin{eqnarray}
K(z)&=&C_{0}(z)+{\cal O}(z^{-2}),\ z\to z_0 \label{con1}\\
K(z)&=&C_m(z)+{\cal O}(1), \ \ \ z\to z_m. \label{con2}
\end{eqnarray}
The number of independent constraints is $(N+2)r-1$ because
${\rm Tr} \ K(z)={\rm Tr} \ L(z)$ is a meromorphic function of $z$.
Thus $\dim \M=(\deg D)r(r-1)-2r+r^2+1$.
The restriction of  $\delta K$ to $\M$ is regular at the poles of $L$ and
has a zero of order 2 at $z_{0}$. Therefore, the form $\omega$
restricted to $\M$ is independent on the choice of the normalization
of the eigenvectors.

\medskip

We can now define the phase space for our system, over which the
form $\omega$ will be intrinsic and non-degenerate. 
The space $\L(D)$ and its subspaces $\M^C$ are invariant under 
the adjoint action $L\to gLg^{-1}$ of $SL_r$. Let 
\be
{\cal P}={\cal P}^C=\M^C/SL_r
\ee 
be the quotient space.
Its dimension equals $\dim {\cal P}=(\deg D)r(r-1)-2r+2$.
Then we have

\bigskip
\noindent
{\bf Theorem 4.} {\it {\rm (a)}
The two-form $\omega$ defined by (\ref{form1}) restricted to $\M$ is gauge invariant
and descends to a symplectic form on ${\cal P}$;\hfill\break
{\rm (b)} The Lax equation (\ref{Lax1}) is Hamiltonian with
respect to $\omega$. The Hamiltonian is 
\beq
\label{H}
H_{n,p}=-{1\over (n+1)}{\rm Tr}\ L^{n+1}(p).
\eeq
{\rm (c)} All the Hamiltonians $H_{n,p}$ are in involution
with respect to $\omega$}.

\bigskip

This provides a straightforward way of exhibiting Lax equations
as Hamiltonian systems.

\subsection{The universal symplectic form: logarithmic version}

There are situations, such as chain models, where a modified version
of the above symplectic form (\ref{symplfo}), defined on a slightly
different phase space, can also be constructed. Which symplectic
form is more appropriate for a given model can be subtle. It can be 
traced back to the fact that there are two basic algebraic structures on
a space of {\it operators}. The first one is the Lie algebra structure defined by the commutator of operators. The second one is the Lie
group structure.
The basic symplectic form introduced 
in the previous section is related to the Lie algebra structure.
We present now a construction of another symplectic structure,
related to the Lie group structure, defined on suitable leaves in $\L(D)$.

\medskip

Consider the open subspace of $\L(D)$ consisting of meromorphic matrix functions which are invertible at a generic point $z$, i.e. 
the subspace of matrices
$L(z)\in \L(D)$ such that $L^{-1}(z)$ is also a meromorphic function. 
We define subspaces of $\L(D)$ with {\it fixed} divisor 
for the poles of $L^{-1}(z)$ as follows.
Fix a set $D^-$ of $(\deg D)r$ distinct points $z_s^-$ and define a
subspace $\L(D,D_-)\subset \L(D)$ by the constraints
\beq
L(z)\in \L(D,D_-):\det L(z)=c{\prod_{s=1}^{Nr} (z-z_s^-)\over \prod_{m=1}^n(z-z_m)^r},
\  c=const\neq 0.
\eeq
If $C_0(z)$ is the same as in (\ref{C}), a subspace $\M_1=\M_1^{C_0}
\subset \L(D,D_-)$
can be defined by the constraints (\ref{con1}).
The following two-form on $\M_1$ is obviously a group version
of (\ref{form1})
\beq\label{gform}
\omega^{\#}={1\over 2} \sum{\rm Res}_{z_a}{\rm Tr}
\left(\Psi^{-1} L^{-1}(z)\delta L(z)\wedge \delta \Psi(z)\right) \,
{dz}
\eeq
Here the sum is taken over all the punctures $z_a=\{z_0,z_m,z_s^{-}\}$.
The subspace $\M_1$ is invariant under the flows defined by the same
Lax equations (\ref{Lax1}), which are also gauge invariant and therefore define flows on the quotient space ${\cal P}_1={\cal P}_1^{C_0}=\M_1^{C_0}/SL_r$.

\bigskip
\noindent
{\bf Theorem 5.} {\it The two-form $\omega^{\#}$ restricted to $\M_1$ is independent
on the normalization of the eigenvectors. It is gauge invariant and descents to a
symplectic form on ${\cal P}_1$. The Lax equation (\ref{Lax1}) is Hamiltonian with
respect to $\omega^{\#}$. The Hamiltonian is}
\beq
H_{n-1,p}=-{1\over n}{\rm Tr}\ L^{n}(p).
\eeq
{\it All the Hamiltonians $H_{n,p}$ are in involution with respect to $\omega^{\#}$}.

\medskip

Thus Theorems 4 and 5 provide a framework for the existence of
so-called bi-Hamiltonian structures.
It was first observed by Magri that the KdV hierarchy possesses
a bi-Hamiltonian structure, in the sense that all the flows of the hierarchy are Hamiltonian
with respect to two different symplectic structures. If $H_n$ is the Hamiltonian
generating the $n$-th flow of the KdV hierarchy 
with respect to the first
Gardner-Zakharov-Faddeev symplectic form, 
then the same flow is generated by the Hamiltonian $H_{n-1}$
with respect to the second
Lenard-Magri symplectic form.

\bigskip

The two symplectic structures $\omega$ and $\omega^{\#}$
are equally good in the case of a single matrix
function $L(z)$, but there is a marked difference between them 
when periodic chains
of operators are considered. Let $L_n(z)\in \L(D)$ be a periodic 
chain of matrix-valued functions with a pole divisor $D$,
$L_n=L_{n+N}$.
The total space of such chains is $\L(D)^{\otimes N}$. The monodromy matrix
\beq
T_n(z)=L_{n+N-1}(z)L_{n+N-2}(z)\cdots L_{n}(z)
\eeq
is a meromorphic matrix function with poles of order $Nh_m$ at the puncture $z_m$,
i.e. $T_n(z)\in \L(ND)$. For different $n$ they are conjugated to each other.
Thus the map
\beq
\L(D)^{\otimes N}\longmapsto \L(ND)/SL_r
\eeq
is well-defined. However,
the natural attempt to obtain a symplectic structure on the space
$\L(D)^{\otimes N}$ by pulling back the first 
symplectic form $\omega$ on
$\L(ND)$ runs immediately into obstacles. The main obstacle is
that the form $\omega$ is only well-defined on 
the symplectic leaves of $\L(ND)$
consisting of matrices with fixed 
singular parts for the eigenvalues at the punctures.
These constraints are {\it non-local},  and cannot be described
in terms of constraints for each matrix $L_n(z)$ separately.

On the other hand, the second symplectic form $\omega^{\#}$
has essentially the desired local property. Indeed,
let $L_n$ be a chain of matrices such that $L_n\in \L(D,D_-)$.
Then the monodromy matrix defines a map
\beq\label{t}
\widehat T:\L(D,D_-)^{\otimes N}\longmapsto \L(ND,\ ND_-)/SL_r
\eeq
The group $SL_r^N$ of $z$-independent matrices $g_n\in SL_r, g_n=g_{n+N}$
acts on $\L(D,D_-)^{\otimes N}$ by the gauge transformation
\beq\label{g}
L_n\to g_{n+1}L_ng_n^{-1}
\eeq
which is compatible with the monodromy matrix map (\ref{t}).
Let the space ${\cal P}_{chain}$ be defined as the corresponding 
quotient space
of a preimage under $\widehat T$ of a symplectic leaf $\P_1^{C_0}\subset \L(ND,ND_-)/SL_r$
\beq
{\cal P}_{chain}=\left(\widehat T^{-1}\left({\cal P}_1^{C_{0}}\right)\right)
/SL_r^N
\eeq
The dimension of this space is equal to
\beq
\dim {\cal P}_{chain}=N(\deg D)r(r-1)-2r+2.
\eeq

\bigskip
\noindent
{\bf Theorem 6.} {\it
The pull-back
by $\widehat T$ of the second symplectic form $\omega_1$
\beq
\omega_{chain}=\widehat T^*(\omega_2)
\eeq
restricted to $\widehat T^{-1}\left({\cal P}_1^{C_{0}}\right)$ is gauge
invariant and descends to a symplectic form on ${\cal P}_{chain}$.
It can also be expressed by the local expression
\beq
\omega_{chain}
=
{1\over 2} \sum{\rm Res}_{z_a}\sum_{n=1}^N{\rm Tr}
\left(\Psi_{n+1}^{-1}\delta L_n(z)\wedge \delta \Psi_n(z)\right)dz
\eeq
where
\beq
\Psi_{n+1}=L_n\Psi_n,\ \ \Psi_{n+N}=\Psi_n K, \ \ K^{ij}=diag(k_i)\delta^{ij},
\eeq
All the coefficients of the characteristic polynomial of $T(z)$ are
in involution with respect to this symplectic form. The number of independent integrals
equals $\dim {\cal P}_{chain}/2$.}

\bigskip
The symplectic forms (\ref{oasym}) and (\ref{osym}) for the spin chain models
corresponding to the $SU(N)$ gauge theory with matter in the
symmetric and the antisymmetric representations can be recognized
as examples of this construction. So is of course the symplectic
structure for the Toda chain.

\subsection{Vector bundles and Lax equations 
on algebraic curves}

As we have seen in the case of the elliptic Calogero-Moser systems,
the spectral parameter of the Lax pair is sometimes defined on an elliptic curve. Below we present briefly results of \cite{kr01a}, where it was
shown that the scheme presented above can be extended to the case of the Lax equations on an arbitrary algebraic curve.

\medskip
The Riemann-Roch theorem shows that the naive direct generalization
of the zero curvature equation for matrix functions which are meromorphic
on an algebraic curve of genus $g>0$ leads to an overdetermined system
of equations. Indeed, the dimension of $r\times r$ matrix functions
of fixed degree $d$ divisor of poles in general position is
$r^2(d-g+1)$. If the divisors of $L$ and $M$ have degrees $n$ and $m$,
then the commutator $[L,M]$ is of degree $n+m$. Thus the number
of equations $r^2(n+m-g+1)$ exceeds the number $r^2(n+m-2g+1)$
of unknown functions modulo gauge equivalence.
\medskip
There are two ways to overcome this difficulty in defining zero curvature
equations on algebraic curves. The first way is based on a choice of $L$
with essential singularity at some point or with entries
as sections of some bundle over the curve. We have seen above that
the standard Lax pair for the elliptic Calogero-Moser system falls
into this category. The second way, based on a theory of {\it high rank}
solutions of the Kadomtsev-Petviashvili equation, was discovered in
\cite{novikov}. There it was shown that if in addition to fixed poles
the matrix functions $L$ and $M$ have $rg$ moving poles of a special
form, then the Lax equation is a well-defined system on the space
of singular parts of $L$ and $M$ at fixed poles.

\medskip
We begin by describing a suitable space of meromorphic matrix functions $L(z)$ on an algebraic curve $\Ga$ of genus $g$.
As before, we fix a divisor $D=\sum h_m z_m$ on $\Ga$, and introduce a
space $\L(D)=\L(D,\Ga)$ of meromorphic matrix functions $L$ on $\Ga$
with a pole of order $h_m$ at $z_m$, such that outside of $D$
$L\in \L(D)$ has simple poles at a set of $rg$ distinct points $\g_s$.
The Laurent expansion of $L$ in the neighborhood of $\g_s$
\beq
\label{Ls}
L={L_{s0}\over z-z_s}+L_{s1}+O(z-z_s),
\ \ z_s=z(\g_s),
\eeq
is assumed to satisfy the following constraints:

(i) the singular term $L_0$ is a {\it traceless, rank 1}  matrix, i.e. it can be represented in the form
\beq\label{Ls0}
L_{s0}=\b_s\a_s^T, \ \ \a_s^T\b_s={\rm tr}\ L_{s0}=0;
\eeq
where $\a_s,\b_s$ are $r$-dimensional vectors;

(ii) $\a_s^T$ is a left eigenvector of the matrix $L_{s1}$
\beq\label{Ls1}
\a_s^TL_{s1}=\a_s^T\kappa_s.
\eeq

The following characterization of the
constraints (\ref{Ls0},\ref{Ls1}) is 
key:
A meromorphic matrix-function $L$ in the neighborhood $U$ of $\g_s$
with a pole at $\g_s$ satisfies the constraints (\ref{Ls0}) and (\ref{Ls1})
if and only if it is of the form
\beq\label{lgauge}
L=\Phi_s(z)\widetilde L_s(z)\Phi_s^{-1}(z),
\eeq
where $\widetilde L_s$ and $\Phi_s$ are holomorphic in $U$, and
$\det \Phi_s$ has at most simple zero at $\g_s$.

\medskip

We would like to emphasize that the points $\g_s$ are not fixed and 
are themselves dynamical variables.
For a non-special degree $N\geq g$ divisor $D$, the space $\L(D)$ is of dimension
\beq\label{dim1}
\dim\  \L(D)=r^2(N+1)\ .
\eeq
As in the genus $g=0$ (rational) case, we define a two-form $\omega$ on $\L(D)$ by the formula
\begin{equation}
\label{formg}
\omega={1\over 2} \left[\sum_{m}{\rm Res}_{z_m}{\rm Tr}
\left(\Psi^{-1} \delta L(z)\wedge \delta \Psi(z)\right) +
\sum_{s}{\rm Res}_{\g_s}{\rm Tr}
\left(\Psi^{-1} \delta L(z)\wedge \delta \Psi(z)\right)\right]\,{dz}
\end{equation}
Here $dz$ is a holomorphic differential on $\Ga$. Here we follow 
again the rule of summing over
all poles of $L$.

\medskip

Let $k_i(z)$ be eigenvalues of $L(z)$, i.e., the roots of the
characteristic equation
\beq\label{char}
\det\ (kI-L(z))=0
\eeq
Then we obtain as previously a foliation structure
on $\L(D)$ by the constraint:

\medskip

{\it The differentials $\delta k_i(z)dz$ are holomorphic in the neighborhood
of all the punctures $z_m$.}

\medskip
In other words, a leaf $\M$ of the foliation is fixed by 
the singular parts of eigenvalues
of $L$ at the points $z_m$ where $dz(z_m)\neq 0$.

\bigskip
\noindent
{\bf Theorem 7.} \cite{kr01a}\ {\it If the divisor $D$ contains the zero divisor
$\K$ of the holomorphic differential $dz$, then the two-form $\omega$ defined
by (\ref{formg}) is invariant under gauge transformations and descends to
a symplectic form on the quotient space ${\cal P}=\M/SL_r$.
The functions $H_{n,p}$ given by (\ref{H}) are in involution with respect to the symplectic form $\omega$}.

\bigskip
\noindent
{\bf Example 1.} Let $D=K$. Then the constraints defining $\M$ are trivial
and $\M=\L(K)$. As shown in \cite{kr01a}, the corresponding 
quotient space
$\L(K)$ is isomorphic to an open subspace of the cotangent bundle
$T^*(Vect)$ of a moduli space of stable, rank r, and degree $rg$ holomorphic vector bundles over $\Ga$. The symplectic form $\omega$ coincides with a canonical symplectic structure on the cotangent bundle.
The integrable structure
of $T^*(Vect)$ was established first by Hitchin \cite{hitchin}
using
a completely different approach.

\bigskip
\noindent
{\bf Example 2.}
Consider the Lax matrices
on an elliptic curve $\G=C/\{2\omega_1{\bf Z}+2\omega_2{\bf Z}\}$ with one puncture,
which we can put at $z=0$ without loss of generality.
In this example we denote the parameters $\g_s$ and $\kappa_s$ by $q_s$ and $p_s$, respectively.
In the gauge $\a_s=e_s,\ e_s^j=\delta_s^j$, the $j$-th column of
the Lax matrix $L^{ij}$ has poles only at the points $q_j$ and $z=0$.
>From (\ref{Ls0}) it follows that $L^{jj}$ is regular everywhere, i.e. it
is a constant. Equation (\ref{Ls1}) implies that $L^{ji}(q_j)=0,\ i\neq j$ and $L^{jj}=p_j$. An elliptic function with two poles and one zero
fixed is uniquely defined up to a constant. It can be written in terms
of the Weierstrass $\s$-function as follows
\beq\label{W}
 L^{ij}(z)=f^{ij}{\s(z+q_i-q_j)\, \s(z-q_i) \s(q_j) \over
\s(z)\s(z-q_j)\,\s(q_i-q_j)\,\s(q_i)}, \ i\neq j;\ \ L^{ii}=p_i.
\eeq
Let $f^{ij}$ be a rank 1 matrix $f^{ij}=a^ib^j$. The equations $\a_i=e_i$ fix the
gauge up to transformations by diagonal matrices. We can use these transformation
to make $a^i=b^i$. The corresponding momentum is given then by the collection
$(a^i)^2$ and we fix it to the values $(a^i)^2=1$.
We compare now the different formalisms for the Lax operator of
the elliptic Calogero-Moser system. In \S 3.2, the Lax operator
$L(z)$ had entries with essential singularities. Its gauge-transform
$\tilde L(z)$ by $G_{ij}=\delta_{ij}e^{q_i\zeta(z)}$ has meromorphic
entries with poles only at the fixed puncture $z=0$, but these entries
are sections of a non-trivial bundle over the elliptic curve.
The present Lax pair $L(z)$ is yet another gauge-transform of the
Lax pair of \S 3.2, this time by the gauge-transformation
\be
G_{ij}=\delta_{ij}\Phi(q_i)
\ee
As we have seen, its entries are now just meromorphic functions,
but with poles at the points $q_j$ as well as at the puncture $z=0$.
 
The new gauge in which the Lax matrix for the elliptic Calogero-Moser system is meromorphic gives a new geometric interpretation of
the elliptic Calogero-Moser system. The dynamical variables $q_i$
can be identified with the so-called Tyurin parameters of the semi-stable
vector bundle over the elliptic curve.

\subsection{$2+1$ soliton equations of zero curvature form}

The integrable models corresponding to the Seiberg-Witten
solutions of gauge theories are mechanical systems with a finite number of
degrees of freedom, and their symplectic
forms are finite dimensional. Nevertheless, the formula \ref{symplfo} is
quite general. It turns out
that it provides also infinite-dimensional symplectic forms for
general soliton equations
admitting a Lax or zero curvature representation. These
symplectic forms are defined
on spaces of functions satisfying suitable constraints, which can
be identified by
a straightforward algorithm \cite{kp98}. This is a welcome
development, since except for the
$R$-matrix approach pioneered by Faddeev and Takhtajan \cite{faddeev}
(see also the recent proposal in \cite{talon}),
there was no systematic
Hamiltonian theory for soliton equations. Although a Hamiltonian
structure was known for
most of them, they did not appear to fit in any
particular scheme.

\medskip

To be specific, we shall construct a whole class of soliton
equations
in zero curvature form and their symplectic forms. Fix an integer
$n\geq 2$, and consider
the space ${\cal M}=\{(u_0,\cdots,u_{n-2});
u_i(x+1,y)=u_i(x,y+1)=u_i(x,y)\}$ of smooth doubly periodic
functions. The points of ${\cal M}$ can be identified with the Lax
operator
\be
L=\p_x^n+\sum_{i=0}^{n-2}u_i\p_x^i
\ee
The integer $n$ classifies different hierarchies of soliton
equations. For example, the case
$n=2$ corresponds to the Kadomtsev-Petviashvili hierarchy, which
reduces to the KdV
hierarchy when $L$ is independent of $y$. The Baker-Akhiezer
function $\psi(x,y,k)$
is defined to be the unique function of the form
\be
\psi(x,y,k)=e^{kx+k^ny+\sum_{i=0}^{n-2}B_i(y)k^i}(1+\xi_1k^{-1}+\xi_2k^{-2}+\cdots)
\ee
characterized by the normalization condition $\psi(0,0,k)=1$ and
\bea
&&(\p_y-L)\psi=0
\nonumber\\
&&\psi(x+1,y,k)=e^k\psi(x,y,k),\ \ \
\psi(x,y+1,k)=e^{K^n}\psi(x,y,k)
\eea
This key observation is that this
determines uniquely both the functions $B_i(y)$ and the coefficients
$\xi_s(x,y)$ as
certain integro-differential expressions
in $(u_0,\cdots,u_{n-2})$, which 
can be written down explicitly.
The dual Baker-Akhiezer function $\psi^{\dagger}(x,ymk)$
is characterized by the condition that it be of the form
\be
\psi^{\dagger}(x,y,k)
=
e^{-(kx+k^ny+\sum_{i=0}^{n-2}B_i(y)k^i)}(1+\xi_1^{\dagger}k^{-1}+\xi_2^{\dagger}k^{-2}+\cdots)
\ee
and satisfies
\be
{\rm Res}_{\infty}\langle \psi^{\dagger}\p_x^m\psi\rangle \,dk=0,
\ \ m=0,1,2,\cdots
\ee
We can now define the phase space of the flow to be the subspace
${\cal M}(b)$ defined by
the equations
\be
{dB_i(y)\over dy}=b_i,\ \ i=0,\cdots,n-2
\ee
where $b_0,\cdots,b_{n-2}$ are some fixed constants. For each $m$,
there is a unique operator
$A_m=\p_x^m+\sum_{i=0}^{m-1}v_{mi}\p_x^i$ satisfying
\be
(A_m-K^m)\psi={\cal O}(k^{-1})\psi
\ee
The soliton equations we consider are the flows on ${\cal L}(b)$
defined by
\be
{\p L\over \p t_m}={\p A_m\over \p y}+[A_m,L]
\ee
where $t_m$ is the time variable of the $m$-th flow. By
construction, these equations
are in zero curvature form. Then

\medskip
\noindent
{\bf Theorem 8.} \cite{kp98} {\it The above flows form an infinite
system of commuting flows on the
space ${\cal L}(b)$. The expression
\be
\label{symplso}
\omega={1\over 2}
{\rm Res}_{\infty}\langle\psi^{\dagger}\delta L\wedge
\delta\psi\rangle\,dk
\ee
defines a symplectic form on ${\cal L}(b)$. With respect to this
form,
the flows are Hamiltonian, with Hamiltonian $nH_{m+n}$, where
$H_s$ is defined by}
\be
k=K+\sum_{s=1}^{\infty}H_sK^{-s}
\ee

\bigskip
It may be helpful to note that, in the context of KdV and the
Kadomtsev-Petviashvili hierarchy, the parameter $k$ above plays
the role of spectral parameter. 
The role of the spectral curve is assumed here by the complex
plane, with
$\infty$ as the distinguished point.Indeed, the Baker-Akhiezer
function $\psi(x,y,k)$ satisfies by definition
the boundary condition
$\psi(x+1,y,k)=e^k\psi(x,y,k)$, and the Lax operator
$L(x)$ acting on such a space should be viewed as dependent
on $k$. Thus the symplectic form in Theorem 8 is a close analogue
of the symplectic forms constructed earlier for Lax operators
with spectral parameter.

\bigskip 

We illustrate this construction when $n=2$ and when $n=3$. The
case $n=2$ corresponds
to the Kadomtsev-Petviashvili hierarchy, and the operator $L$ is
given by $L=\p_x^2+u$.
A straightforward calculation shows that
\be
\psi(x,y,k)=e^{kx+k^2y+B(y)}(1+\xi_1k^{-1}+\cdots)
\ee
with
$B(y)=\int_0^y\int_0^1 u(x,\alpha)dxd\alpha$, and
$\p_x\xi_1=-{1\over 2}u+{dB\over dy}$.
The space ${\cal L}(b)$ is defined then by the constraint
\be
\int_0^1u(x,y)dx=b
\ee
and we have $\xi_1=-{1\over 2}\int^xu+bx$. It follows readily that
$\psi^{\dagger}=e^{-(kx+k^2y+B(y))}(1-\xi_1k^{-1}+\cdots)$, and
the symplectic form
becomes
\be
\omega={\rm Res}_{\infty}\langle (1-\xi_1k^{-1}+\cdots)\,\delta
u\wedge
(\delta\xi_1k^{-1}+\cdots)\rangle \,dk
=\langle \delta u\wedge \delta\xi_1\rangle
=-{1\over 2}\langle\delta u(x)\wedge \int^x\delta u\rangle
\ee
This is the Gardner-Faddeev-Zakharov symplectic form for the KdV
equation.

The case $n=3$ corresponds to the Boussinesq hierarchy.
Here $L=\p_x^3+u\p_x+v$, and the space ${\cal L}(b_0,b_1)$ is the
space of doubly
periodic functions $u$ and $v$ satisfying
\be
\int_0^1 u(x,y)dx=b_0,\ \ \int_0^1v(x,y)=b_1.
\ee
We find the following expression for the symplectic form
\be
\omega=-{1\over 3}(\delta u\wedge \int_{x_0}^x\delta v dx+\delta
v\wedge\int_{x_0}^x\delta u dx)
\ee

\subsection{Calogero-Moser field equations and isomonodromy equations}

Suitably interpreted, the universal formula \ref{symplfo} can produce a
Hamiltonian structure for differential equations in diverse contexts, including field versions of Calogero-Moser equations and monodromy
equations \cite{kr01a} \cite{kr01b}.
We describe these symplectic structures briefly.

\medskip

The field version of Calogero-Moser system
arises as a zero curvature equation for a Lax pair
$L(x,q)$, $M(x,q)$ 
\be
\label{cmfield}
L_t=M_x+[M,L]
\ee
with elliptic spectral parameter $q$.
Here $L(x,q)$ is a $r\times r$ matrix, periodic of period $1$ in
$x$ and meromorphic in $q$. The set-up is analogous to the one
considered in Example 2 in the previous section, with an additional
space variable $x$. The matrix $L^{ij}(x,q)$ now has poles in $q$
only at $q=0$ and $q=q_j(x)$. On a suitable space $\L$ of operators
$L(x,q)$ satisfying some specific constraints on their poles and
residues, one can construct as before an infinite number of $r\times r$
matrix valued operators $M_n(x,q)$ whose coefficients are differential
expressions in terms of the coefficients of $L(x,q)$, which satisfy
the condition that $\p_xM+[M,L]$ is tangent to $\L$.
The equation (\ref{cmfield}) defines then an infinite number
of commuting flows on $\L$. As in the construction of the symplectic
form for the Kadomtsev-Petviashvili hierarchy, a unique suitably normalized solution
$\Psi(x,q)$ of the equation $(\p_x-L(x,q))\Psi(x,q)=0$
can be found, with monodromy of the form
\be
\Psi(x+1,q)=e^{p(q)}\Psi(x,q)
\ee
for some unique $p(q)$. Then the following modification
of the universal symplectic form (\ref{symplfo}) 
\be
\omega=
-{1\over 2}\sum_{\alpha}{\rm
Res}_{\alpha}Tr(\int_{x_0}^{x_0+1}(\Psi^{-1}\delta L\wedge
\delta\Psi)dx
-(\Psi^{-1}\delta\Psi)(x_0)\wedge \delta p)dz
\ee
defines a symplectic structure on $\L$ with respect to which
all the flows (\ref{cmfield}) are Hamiltonian. For $r=2$, one
finds the following expression for the corresponding Hamiltonian
\be
H=\int \big(p^2(1-q_x^2)-{q_{xx}^2\over 2(1-q_x)^2}
+2(1-3q_x^2)\,\wp(2q)\big)dx
\ee
giving the equations of motion for the poles $q(x)$. In terms of $q(x)$
and its conjugate variable $p(x)$, the symplectic
structure leads to the simple Poisson bracket
\be
\{p(x),q(y)\}=\delta(x-y).
\ee
It is noteworthy that after a change of variables, this system
coincides with the well-known Landau-Lifshitz equation.

\medskip

Next, we discuss the isomonodromy problem. Let
$V$ be a stable rank $r$ and degree $rg$
holomorphic bundle on a genus $g$ Riemann surface
$\Gamma$, and let
$[V]=\{\gamma_s\}$ be the divisor of its determinant bundle.
Let $D=\sum_m(h_m+1)P_m$ be a divisor
which does not intersect $[V]$.
Consider meromorphic connections
$L=L(z)dz$ on $V$ with poles at $P_m$ of degree $\leq h_m+1$.
As shown in \cite{kr01b},
they can be represented by meromorphic matrix-valued 
differentials $L$ with a pole of order $h_m+1$ at $P_m$,
with the property that outside $D$, the differential
$L$ has simple poles at the points $\gamma_s$.
The Laurent expansion of $L$ in a neighborhood of $\gamma_s$
\be
L=\big({L_{s0}\over z-z_s}+L_{s1}+{\cal O}(z-z_s)\big)dz,
\ \ z_s=z(\gamma_s)
\ee
is assumed to satisfy the following constraints

(i) the singular term $L_{s0}$ is a rank $1$ and trace $1$ matrix, i.e.,
\be
L_{s0}=\beta_s\alpha_s^T,
\ \ \
\alpha_s^T\beta_s=Tr\,L_{s0}\,=1,
\ee
where $\alpha_s,\beta_s$ are $r$-dimensional vectors.

(ii) The vector $\alpha_s^T$ is a left eigenvector of the matrix $L_{s1}$
\be
\alpha_s^TL_{s1}=\alpha_s^T\kappa_s.
\ee

\medskip
The following characterization of these constraints is essential:
A meromorphic matrix-valued function $L$ in a neighborhood $U$
of $\gamma_s$
with pole at $\gamma_s$ satisfies them if and only if it is of the
form
\be
\label{charact}
L=d\Phi_s(z)\Phi_s(z)^{-1}
+
\Phi_s(z)\tilde L_s(z)\Phi_s^{-1}(z),
\ee
where $\tilde L_s$ and $\Phi_s$ are holomorphic in $U$,
and ${\rm det}\,\Phi_s$ has at most a simple zero at $\gamma_s$.
For any point $P\in\Gamma$, it follows from (\ref{charact})
that the equation
\be
\label{baiso}
d\Psi=L\Psi
\ee
admits a multi-valued solution in $\Gamma\setminus D$.
The analytic continuation of $\Psi$ defines the monodromy representation
of $L$. The isomonodromy problem is the problem
of describing the deformations of $L$ which preserve the monodromy
representation as well as certain local data at the singular points
$P_m$ called Stokes matrices and exponents.
In order to define these data, it is necessary to
fix a normalization point $Q\in\Gamma$, and $h_m$-jets
of local coordinates in the neighborhoods of the punctures
$P_m$.

\medskip
Let $h=\{h_m,\sum_m(h_m+1)=N\}$ be a set of non-negative integers.
We denote by ${\cal M}_{g,1}(h)$
the moduli space of smooth genus $g$ algebraic curves
with a puncture $Q\in \Gamma$, and fixed $h_m$-jets $[w_m]$
of local coordinates $w_m$ in the neighborhoods of the punctures
$P_m$. The space ${\cal M}_{g,1}(h)$ has dimension
\be
{\rm dim}\,{cal M}_{g,1}(h)=3g-2+N
\ee
The space ${\cal A}(h)$ of admissible meromorphic
differentials $L$ on algebraic curves with fixed multiplicities
$h_m+1$ at the poles can be viewed as the total
space of the bundle
\be
{\cal A}(h)
\longrightarrow
{\cal M}_{g,1}(h)=\{\Gamma,P_m,[w_m],Q\}
\ee
For a fixed point of ${\cal M}_{g,1}(h)$,
the monodromy data, Stokes matrices, and exponents uniquely
define $L$. In 
\cite{kr01b}, special coordinates $t_a$ were introduced
and it was shown that the isomonodromy deformations on the space
${\cal A}$ of meromorphic connections 
are described by the zero curvature equations
\be
\p_{t_a}\tilde L-dM^{(a)}+[\tilde L,M^{(a)}]=0
\ee
for suitable $M^{(a)}$.
The corresponding flows commute. Furthermore,
as in all the models described previously, they
turn out to be Hamiltonian with respect to the symplectic
structure defined by
\be
\omega=-{1\over 2}\big({\rm Res}_{[V]}Tr (\psi^{-1} \delta 
L\wedge \delta\psi)
+
{\rm Res}_{D}Tr (\Psi_m^{-1} \delta 
L\wedge \delta\Psi_m)
\ee
Here $\psi$ and $\psi_m$ are the solutions of the equation
(\ref{baiso}) in the neighborhoods of the points
in the divisors $[V]$ and $D$ respectively.
An explicit expression of the symplectic form in terms of
the monodromy data and Stokes matrices was also found in
\cite{kr01b}. It should be stressed that,
even in the case of genus $0$, a symplectic structure on
the space of Stokes matrices with one irregular singularity
of order $2$ and one regular singularity was found only recently
\cite{boalch}. In the case of elliptic curves, the monodromy data
consists of just two matrices $A$ and $B$, modulo conjugation,
and the monodromy matrix is $J=B^{-1}A^{-1}BA$.
The symplectic form $\omega$ becomes in this case
\be
\omega(A,B)
=
Tr\big(B^{-1}\delta B\wedge \delta A A^{-1}-A^{-1}\delta A\wedge
\delta BB^{-1}+
\delta J J^{-1}\wedge B^{-1}A^{-1}\delta(AB)\big)
\ee
It had also been found in \cite{goldman} under a different form.

\bigskip
\vfill\break

\end{document}